\documentclass[pra,superscriptaddress,amsmath,amssymb]{revtex4}
\usepackage{graphicx,color,appendix}
\usepackage{dcolumn}
\usepackage{bm}

\newtheorem{theorem}{Theorem}
\newtheorem{remark}[theorem]{Remark}

\def\diag{\rm diag}

\def\Imag{\operatorname{Im}}
\def\Real{\operatorname{Re}}

\begin{document}

\title{Designing quantum memories with embedded control: photonic circuits for autonomous quantum error correction}
\author{Joseph Kerckhoff} \email{jkerc@stanford.edu}
\affiliation{Edward L.\ Ginzton Laboratory, Stanford University, Stanford, California 94305, USA}
\author{Hendra I. Nurdin} \email{Hendra.Nurdin@anu.edu.au}
\affiliation{Edward L.\ Ginzton Laboratory, Stanford University, Stanford, California 94305, USA}
\affiliation{Department of Information Engineering, The Australian National University, Canberra, ACT 0200, Australia}
\author{Dmitri S. Pavlichin} \email{dmitrip@stanford.edu}
\affiliation{Edward L.\ Ginzton Laboratory, Stanford University, Stanford, California 94305, USA}
\email{dmitrip@stanford.edu}
\author{Hideo Mabuchi} \email{hmabuchi@stanford.edu}
\affiliation{Edward L.\ Ginzton Laboratory, Stanford University, Stanford, California 94305, USA}
\email{hmabuchi@stanford.edu}

\date{\today}


\maketitle

\noindent Quantum error correction (QEC)~\cite{Gott09} is fundamental for quantum information processing but entails a substantial overhead of classically-controlled quantum operations, which can be architecturally cumbersome to accommodate. Here we discuss a novel approach to designing elementary QEC memory cells, in which all control operations are performed autonomously by an embedded optical feedback loop. Our approach is natural for nanophotonic implementations in which each qubit can be coupled to its own optical resonator, and our design for a memory cell based on the quantum bit-flip or phase-flip code requires only five qubit-cavities (three for the register and two for the controller) connected by wave-guides. The photonic QEC circuit is entirely on-chip, requiring no external clocking or control, and during steady-state operation would only need to be powered by the injection of constant-amplitude coherent fields.

While our approach exploits rigorous theoretical results derived using techniques from quantum field theory and coherent-feedback quantum control~\cite{Goug09a,BvHJ07,HP84,KRP92,Fagn90,GZ00}, we   devote the main body of this article to an essentially intuitive presentation of our QEC photonic network.  Following this, a more complete modeling of the system is presented in extended appendices.  Appendix \ref{sec:prelim} briefly recounts the Hudson-Parthasarathy stochastic calculus~\cite{BvHJ07,HP84,KRP92} and a modeling of interconnected open Markov quantum components~\cite{Goug09a} that provide a formalism well-suited to modeling complex quantum optical networks.  Appendix \ref{sec:subsystems-derivations} then provides the specific models for several of the components in our network and introduces a key limit theorem~\cite{BvHJ07} necessary to produce a computationally-tractable model for the overall network.  Finally, Appendix~\ref{sec:network} constructs the full network model and derives its dynamics.  

\begin{figure}[b!]
\includegraphics[width=0.48\textwidth]{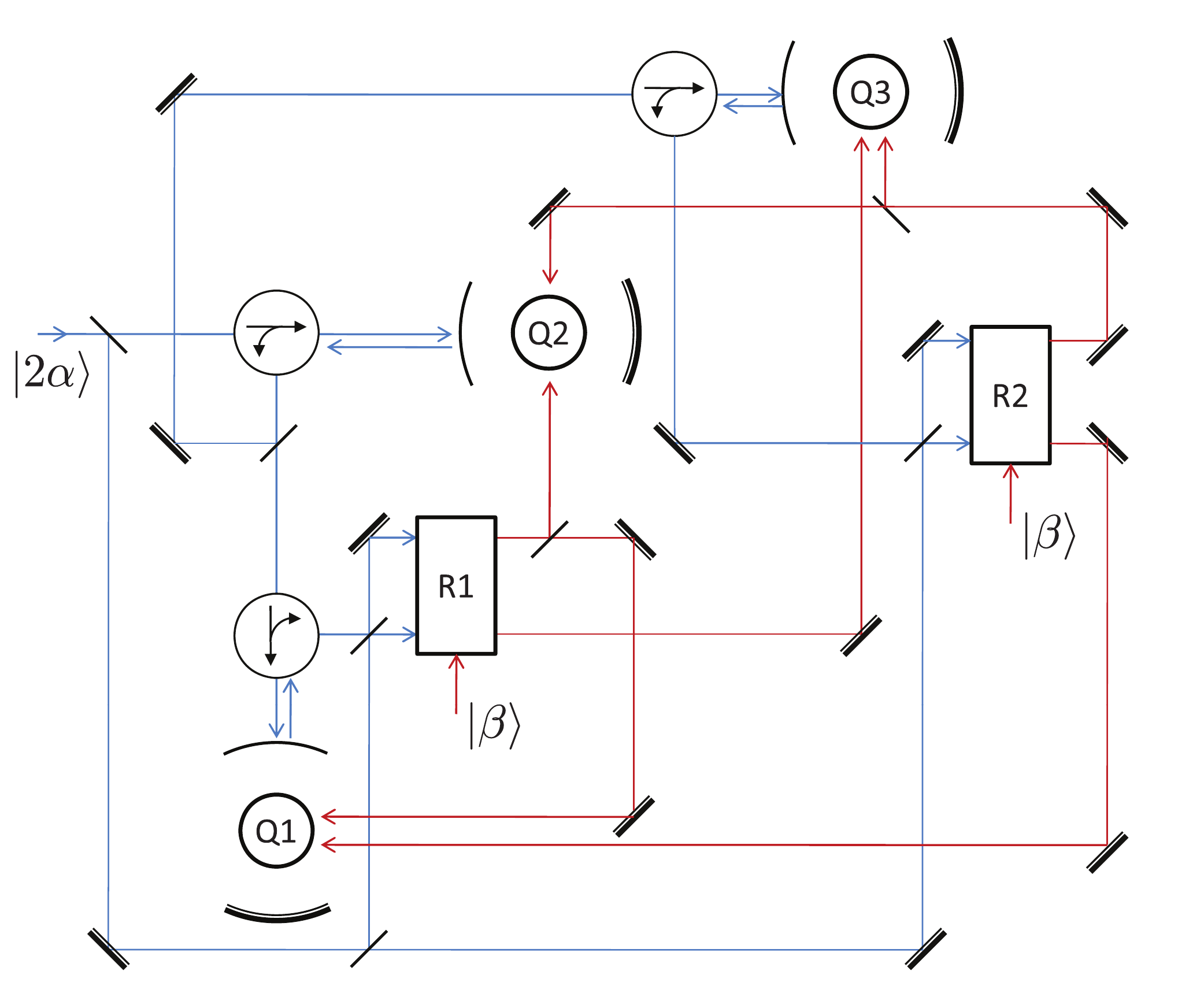}
\caption{\label{fig:relaynet} {\bf Photonic QEC network schematic.}  Schematic diagram of a coherent-feedback quantum memory showing qubits in cavities (Q1, Q2 and Q3), circulators, beam-splitters, steering mirrors and relays (R1 and R2).}
\vspace{-0.1in}
\end{figure}

The photonic circuit shown in Fig.~\ref{fig:relaynet} implements continuous QEC based on the bit-flip or phase-flip code~\cite{Gott09}. Q1, Q2 and Q3 represent the register qubits; the blue signal lines indicate the routing of a laser beam (injected in the coherent state $\vert2\alpha\rangle$, where $\vert\alpha\vert^2$ has units photons time$^{-1}$) used for error detection (following principles described in~\cite{Kerc09a}); $R1$ and $R2$ represent two qubit-based photonic relays (as analyzed in~\cite{Mabu09b}); and the red signal lines indicate pathways for optical feedback using two additional laser beams (injected in the coherent state $\vert\beta\rangle$) for corrective Raman bit- or phase-flips. Such a circuit could be implemented using solid-state qubits embedded in nanophotonic resonators connected by single-mode wave-guides~\cite{Engl07,Barc09}, or using microwave components as in circuit QED~\cite{Maje07}.

%

 The quantitative performance of such a quantum memory will of course depend on physical parameters such as the qubit-cavity coupling strengths and the cavity decay rates, but below and in Appendix~\ref{sec:subsystems-derivations} we show that our approach does not require any fine-tuning of the parameters and requires only that certain ratios be large. It is interesting to note that these ratios become large in a {\em small volume limit} for the optical resonators, which is a natural limit to consider in nanophotonic implementations. Passage to the small volume limit also gives rise to drastic simplifications of the quantum input-output models for the components in the QEC circuit, and thus emerges as a significant new abstraction principle for the analysis of nanophotonic signal-processing networks.

\begin{figure}[tb!]
\includegraphics[width=0.3\textwidth]{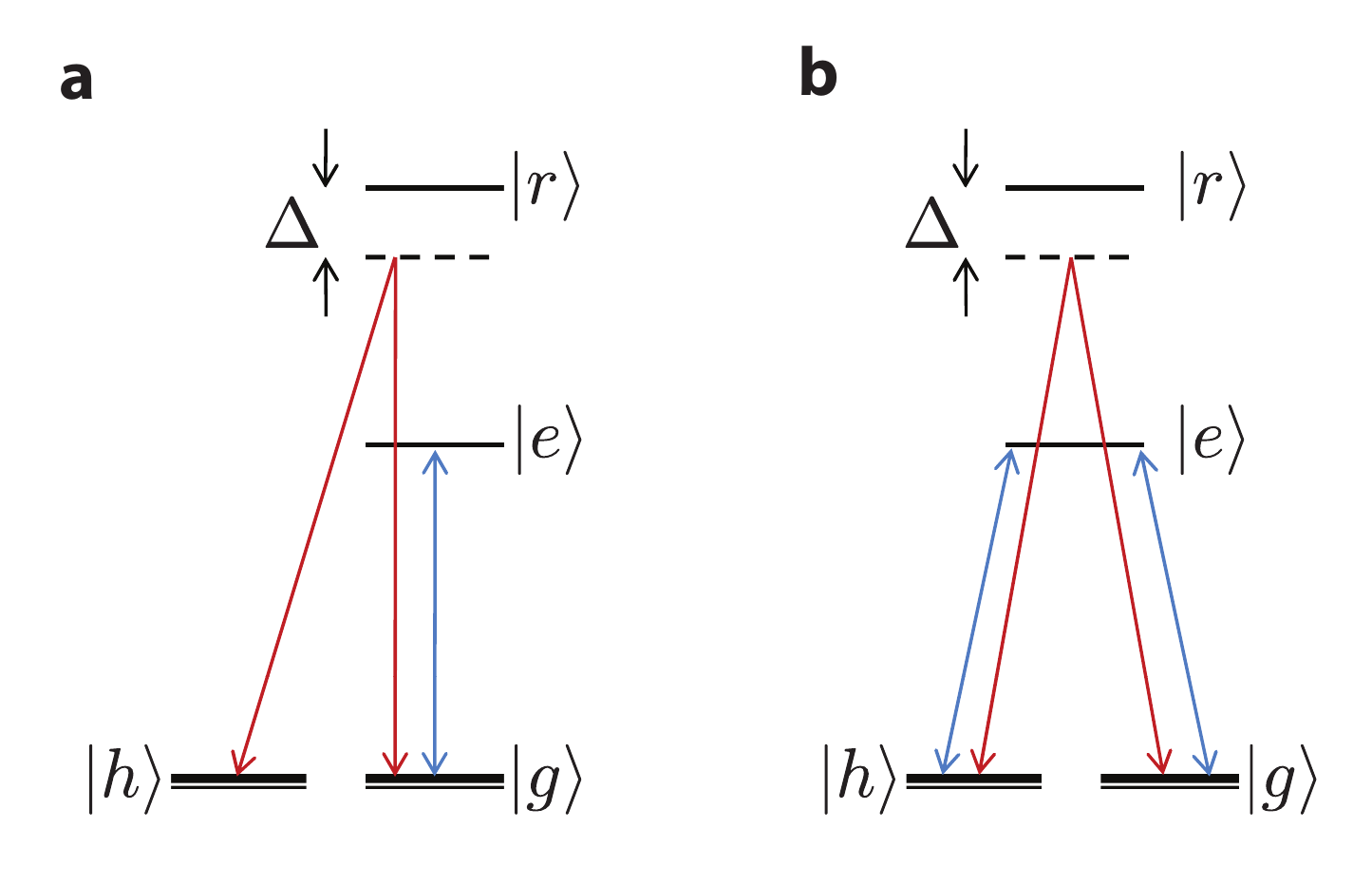}
\caption{\label{fig:levels} {\bf The cQED qubit.}  Level diagrams of {\bf a}, $Z$ probe interaction, {\bf b}, $X$ probe interaction.}
\vspace{-0.1in}
\end{figure}

The physical qubits Q1, Q2 and Q3 are actually multi-level `atoms' or comparable solid-state emitters, {\it e.g.}, nitrogen-vacancy centers. Fig.~\ref{fig:levels} shows the level structure presumed for the individual atoms and the couplings utilized for the bit/phase-flip schemes, with $\{\vert g\rangle,\vert h\rangle\}$ corresponding to the single qubit states $\{\vert 0\rangle,\vert1\rangle\}$. In the bit-flip implementation we assume that each qubit is strongly coupled to a single-mode, single-sided cavity on its $\vert g\rangle\leftrightarrow\vert e\rangle$ transition as discussed in \cite{Kerc09a}. As a result, the phase acquired on reflection by a probe beam is $\pi$ if the qubit is in state $\vert g\rangle$ and $0$ if it is in $\vert h\rangle$. While details of the probe-qubit interaction can be investigated for any desired physical parameters using the familiar cavity QED master equation~\cite{Kerc09a}, for our network calculation we utilize a simple abstraction obtained when the vacuum Rabi frequency $g_p$ and cavity field decay rate $\kappa_p$ are taken to infinity with the decoherence rates held fixed. Because $g_p$ and $\kappa_p$ both increase with decreasing resonator volume, we term this a `small volume' limit. The simplified model is derived using a quantum stochastic differential equation (QSDE) limit theorem~\cite{Bout08} following a similar procedure to that of~\cite{Kerc09a} (see Appendix~\ref{sec:probe-derivation}). In the phase-flip scenario we modify the scheme by assuming a linearly polarized cavity mode (as in a micro-disk or photonic crystal resonator) that couples to both a $\sigma_-$ transition from $\vert g\rangle\leftrightarrow\vert e\rangle$ and a $\sigma_+$ transition from $\vert h\rangle\leftrightarrow\vert e\rangle$.

\begin{figure}[tb!]
\includegraphics[width=0.48\textwidth]{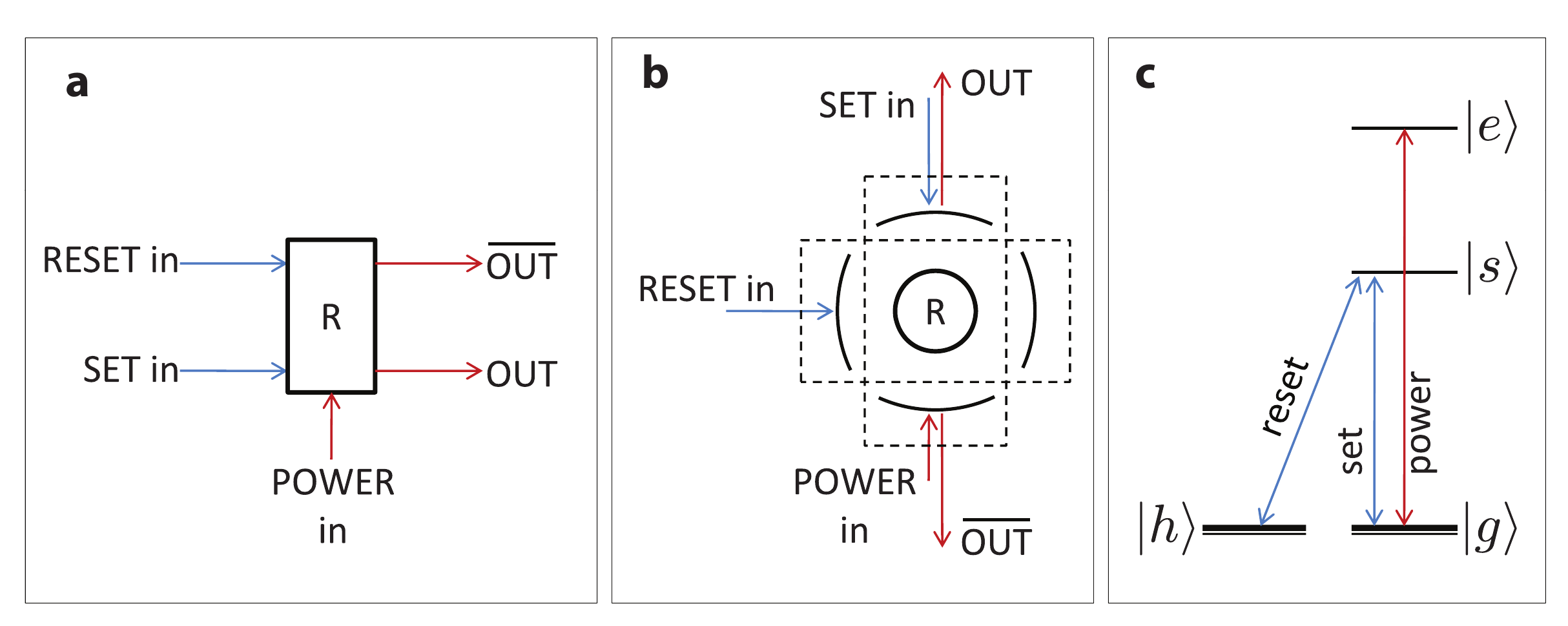}
\caption{\label{fig:relays} {\bf The cQED photonic relay.}  Details of the cQED relay component model (modified from \cite{Mabu09b}): {\bf a}, input and output ports, {\bf b}, coupling of input/output fields to resonant modes of two cavities, and {\bf c}, relay internal level diagram.}
\vspace{-0.1in}
\end{figure}

The central component in the network's QEC `controller' is a recently proposed~\cite{Mabu09b} cavity-QED relay, some details of which are presented in Fig.~\ref{fig:relays}. Three cavity modes are resonant with various transitions in an intra-cavity atom, as shown in Fig.~\ref{fig:relays}~{\bf b} and {\bf c}. In a similar small volume limit \cite{Mabu09b}, the atomic dynamics are limited to only the ground states, and when the atom is in state $\vert h\rangle$ ($\vert g\rangle$) the cavity's POWER input is connected to the OUT ($\overline{\rm OUT}$) output port. A probe signal driving the SET input port (in the absence of signal at the RESET input) causes the relay state to decay to $\vert h\rangle$, while driving only the RESET input induces decay to $\vert g\rangle$. Through these internal dynamics, a probe signal entering either the SET or RESET port of the relay routes the POWER input out either the OUT or $\overline{\rm OUT}$ port, respectively.

We now trace through the circuit dynamics of Fig.~\ref{fig:relaynet} in more detail, assuming a bit-flip code implementation. A coherent input $\vert 2\alpha\rangle$ is split into probe and reference beams. The probe beam first reflects from the cavity containing Q2, gaining a $\pi$ or $0$ phase shift depending on the state of that atom, putting the probe signal into a coherent state with amplitude $\sqrt{2}\alpha Z^{(Q2)}$ (where $Z^{(Q2)}=\vert h\rangle\langle h\vert-\vert g\rangle\langle g\vert$ on Q2 and the identity operation on all other Hilbert spaces is implied---see Appendix \ref{sec:network}). The probe signal is again split and the two resulting fields separately probe the cavities containing Q1 and Q3, resulting in signals with coherent amplitudes $\alpha Z^{(Q1)}Z^{(Q2)}$ and $\alpha Z^{(Q2)}Z^{(Q3)}$. These are interfered with copies of the reference beam, producing four signals with amplitudes $\alpha(I\pm Z^{(Q1)}Z^{(Q2)})/\sqrt2$ and $\alpha(I\pm Z^{(Q2)}Z^{(Q3)})/\sqrt2$. The `+' signals are sent to the SET inputs of relays R1 and R2, the `-' to the RESET inputs. In this way, the SET (RESET) input to R1 receives a coherent input signal if Q1 and Q2 have even (odd) parity, while the RESET (SET) receives only a vacuum input. Similarly for R2 and qubits Q2 and Q3. The POWER inputs to R1 and R2 are two more coherent fields, each with amplitude $\beta$ and frequency and polarization that drive the remaining, far-detuned atomic transitions depicted in Fig.~\ref{fig:levels}. The $\overline{\rm OUT}$ port of R1 (R2) is directed to both Q1 and Q2 (Q2 and Q3). The OUT port of R1 (R2) is directed to Q3 (Q1). When a qubit is simultaneously illuminated by two of these feedback beams, Raman resonance conditions are satisfied, inducing a coherent Rabi oscillation between the states of that single qubit. When only one feedback beam illuminates an atom, the beam is assumed to be sufficiently detuned such that it only imparts an AC Stark shift to one of the atomic ground states. Remarkably, a proper choice of the ground state each feedback beam targets prevents Stark shift-induced dephasing of the stored codeword \emph{in any error state} (see Appendix~\ref{sec:SS}).

Assuming the register qubits are initialized in a proper bit-/phase-flip codeword, this network will self-stabilize against bit/phase-flip errors without any external input other than the coherent probe and Raman fields ($\alpha$, $\beta$, and all Hamiltonian coupling constants are constant in time). For an intuitive understanding of the network, note that if no flip errors occur, the SET inputs to the relays are activated, Q1 and Q3 only receive one feedback signal each and no corrective dynamics occur. If for example Q2 flips, the RESET inputs on both R1 and R2 are activated, Q2 receives both feedback signals, while Q1 and Q3 again get only one each. A corrective flipping Hamiltonian is thus implemented on Q2 until a flip is `complete,' which in turn drives the relay switches back to the `no-error' state, which then extinguishes the flipping Hamiltonian. This coherent feedback network thus implements a continuous version of the usual discrete-step strategy that diagnoses and corrects particular errors without disturbing the information content of the encoded joint state.

After interconnecting simplified (small-volume) models for the bit-flip components according to the bit-flip circuit diagram, including individual bit-flip error processes, and adiabatically eliminating of all but the qubit and relay ground states, we obtain (see Appendix~\ref{sec:network}) the closed-loop master equation
\begin{equation}
\dot{\rho}_t=-i[H,\rho_t] + \sum_{j=1}^{7}\left( L_j\rho_tL_j^* - \frac{1}{2}\{L_j^*L_j,\rho_t\}\right),\label{eq:meq}
\end{equation}
where
\begin{eqnarray}
H&=&\Omega\biggl(\sqrt{2}Z^{(Q1)}\Pi_g^{(R1)}\Pi_h^{(R2)}+X^{(Q2)}\Pi_g^{(R1)}\Pi_g^{(R2)}-\sqrt{2}X^{(Q3)}\Pi_h^{(R1)}\Pi_g^{(R2)}-\nonumber\\
&&\Pi_g^{(R1)}(\Pi_g^{(Q1)}+\Pi_h^{(Q2)})-2\Pi_h^{(R1)}\Pi_g^{(Q3)}-\Pi_g^{(R2)}(\Pi_g^{(Q2)}+\Pi_h^{(Q3)})-2\Pi_h^{(R2)}\Pi_h^{(Q1)}
\biggr),\nonumber\\
L_1 &=& \frac{\alpha}{\sqrt{2}}\left\{\sigma_{hg}^{(R_1)}(1+Z^{(Q1)}Z^{(Q2)})-\Pi_g^{(R_1)}(1-Z^{(Q1)}Z^{(Q2)})\right\},\nonumber\\
L_2 &=& \frac{\alpha}{\sqrt{2}}\left\{\sigma_{gh}^{(R_1)}(1-Z^{(Q1)}Z^{(Q2)})-\Pi_h^{(R_1)}(1+Z^{(Q1)}Z^{(Q2)})\right\},\nonumber\\
L_3 &=& \frac{\alpha}{\sqrt{2}}\left\{\sigma_{hg}^{(R_2)}(1+Z^{(Q3)}Z^{(Q2)})-\Pi_g^{(R_2)}(1-Z^{(Q3)}Z^{(Q2)})\right\},\nonumber\\
L_4 &=& \frac{\alpha}{\sqrt{2}}\left\{\sigma_{gh}^{(R_2)}(1-Z^{(Q3)}Z^{(Q2)})-\Pi_h^{(R_2)}(1+Z^{(Q3)}Z^{(Q2)})\right\},\nonumber\\
L_5 &=& \sqrt{\Gamma}X^{(Q1)},\quad L_{6} = \sqrt{\Gamma}X^{(Q2)},\quad L_{7} = \sqrt{\Gamma}X^{(Q3)}.\label{eq:Ldefs}
\end{eqnarray}
where $X^{(C)} = \vert h\rangle\langle g\vert+ \vert g\rangle\langle h\vert$,  $\sigma_{jk}^{(C)} =  \vert j\rangle\langle k\vert$ and $\Pi_j^{(C)}= \vert j\rangle\langle j\vert$ on component $C$, each qubit suffers random flips at a rate $\Gamma$, and the `feedback parameter' $\Omega$ is proportional to $\vert\beta\vert^2/\Delta$ (note that several $L_j$ terms corresponding to strong dephasing of the relay states have been omitted as they have no appreciable effect on our dynamics). An analogous procedure using the phase-flip component models leads to a master equation with the substitutions $Z^{(Qi)}\leftrightarrow X^{(Qi)}$, $2\Pi^{(Qi)}_{^h_g}\rightarrow1\pm X^{(Qi)}$.  While $L_5$ through $L_7$ drive the corrupting bit-flip processes, the first three terms of the Hamiltonian (H) and remaining `collapse operators' work in concert to implement the corrections: $L_1$ through $L_4$ chiefly drive the state of the two relays to reflect the qubit parities, and each bit-flipping term in the Hamiltonian is only active for a particular `syndrome state'~\cite{Gott09}, as indicated by the relays.  The remaining Hamiltonian terms physically correspond to AC Stark shifts, whose effects have been minimized by design (see Appendix~\ref{sec:SS}). Indeed from the general structure of equation~(\ref{eq:Ldefs}) it is straightforward to intuit master equations that correspond abstractly to memories based on other stabilizer codes, highlighting the potential of our approach for developing technology-inspired abstractions in engineering of quantum optical networks.

\begin{figure}[tb!]
\includegraphics[width=0.45\textwidth]{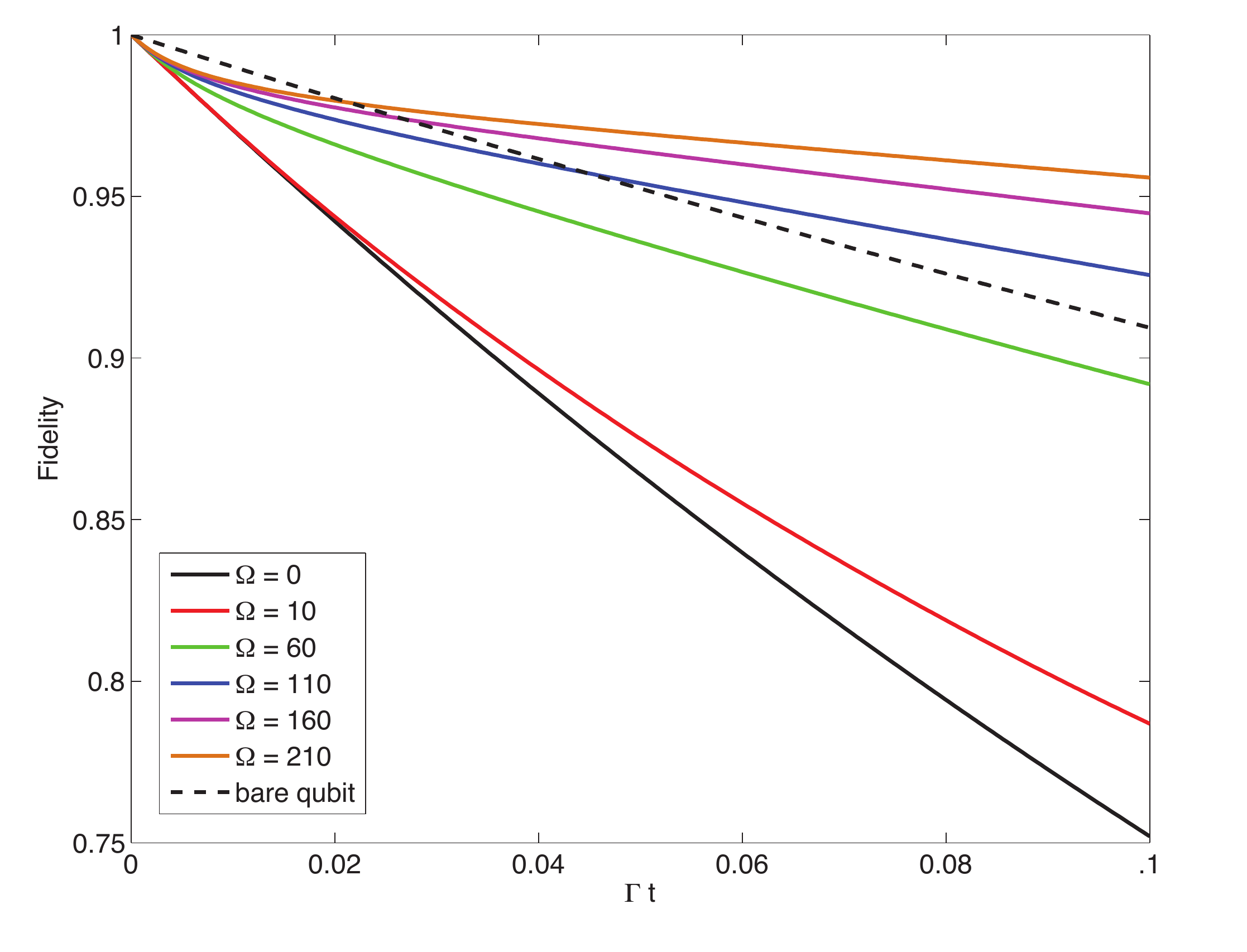}
\caption{\label{fig:meq} {\bf Network performance.} Decay of fidelity, $\langle\Psi_0\vert \rho_t \vert\Psi_0\rangle$, for several values of the feedback parameter $ \Omega=\vert\beta\vert^2\gamma/2\Delta$ (see Appendix~\ref{sec:feedback_network}), $\alpha= \Omega/8$, and $\Gamma=0.1$.  The fidelity decay of a single, bare qubit also suffering bitflip errors at rate $\Gamma=.1$ is also shown.} \vspace{-0.1in} \end{figure}

Note that a network simulation with full physical models of all the components would be beyond our present computational capabilities; judicious use of the small-volume abstraction, which allowed us to derive the easily interpreted closed-loop master equation~(\ref{eq:meq}), has thus been essential in our effort to prove that autonomous QEC with fully embedded control is possible in principle in nanophotonic implementations. Previous studies~\cite{Kerc09a,Mabu09b} have shown that the idealized behaviors are reasonably well approximated in component simulations using realistic physical parameters.

In Fig.~\ref{fig:meq} we display some illustrative numerical integrations~\cite{Tan99} of equation~(\ref{eq:meq}) for the bit-flip scenario. With the flip rate set to $\Gamma=0.1$, the feedback amplitude is varied from $\Omega=0$ (no feedback) to $\Omega=210$, with the probe amplitude kept at $\alpha=\Omega/8$.  The initial state of the qubit register is chosen as $\vert\Psi_0\rangle=(\vert ggg\rangle -i\vert hhh\rangle)/\sqrt{2}$, $\rho_0 = \vert\Psi_0\rangle\langle\Psi_0\vert$, and the fidelity decay $\langle\Psi_0\vert \rho_t\vert\Psi_0\rangle$ is computed to quantify the decoherence suppression achieved by the coherent feedback loop. With $\Omega=0$ the fidelity decay is identical to that of the three-qubit register without any measurement or feedback, while for larger values of $\Omega$ and longer storage times, the fidelity of the encoded qubit is clearly improved relative to what it would be for a bare qubit. Variation in performance for different $\vert\Psi_0\rangle$ is only due to the different susceptibilities of codewords to bit-flip errors. Because of the small, but finite likelihood that additional errors accrue before the correction of a single flip is implemented, fidelity is steadily lost for finite probe and feedback strengths, in analogy to the failure of the discrete time bit-flip code to correct multiple flips.

In conclusion, we have described a novel approach to continuous quantum error correction based on stationary coherent feedback. Our scheme may match well with hardware implementations that feature solid-state qubits embedded in planar circuits of electromagnetic resonators and wave-guides, and establishes important new connections between QEC and coherent-feedback quantum control theory \cite{Jame07a,Mabu08a,Jame10a}. While our current results are limited to the bit/phase-flip scenarios, we hope in future work to find ways to combine bit-flip and phase-flip probe interactions (possibly employing either the five-qubit code~\cite{Gott09} or Bacon-Shor subsystem code~\cite{Alif07}) to design feedback networks that can correct arbitrary single-qubit errors.

\begin{acknowledgments}
This work has been supported by IARPA (W911NF-08-1-0491) and by HP Labs. HN acknowledges the support of the Australian Research Council (DP0986615).
\end{acknowledgments}

\appendix

\section{QSDE formalism}
\label{sec:prelim} To describe our QEC coherent-feedback network, we employ the formalism of
the Hudson-Parthasarathy stochastic calculus and quantum stochastic differential equations (QSDEs), and the concatenation and series product rules for interconnecting open Markov quantum components. Here we shall only give a brief reminder of these formalisms, a detailed exposition and nice introduction to the formalisms, including several illustrative examples, can be found in \cite{Goug09a,BvHJ07}; see also the references cited therein. 

\subsection{Hudson-Parthasarathy QSDEs and unitary evolution of open Markov quantum systems}
\label{sec:HP-QSDEs}
In the following we shall use the notations: $i=\sqrt{-1}$,
$^*$ will denote the adjoint of a linear operator as well as the
conjugate of a complex number, if $A=[a_{jk}]$ is a matrix of
linear operators or complex numbers then $A^{\#}=[a_{jk}^*]$, and
$A^{\dag}$ is defined as $A^{\dag}=(A^{\#})^T$, where $^T$ denotes
matrix transposition. We also define $\Real\{A\}=(A+A^{\#})/2$ and
$\Imag\{A\}=(A-A^{\#})/2i$ and denote the identity matrix by $I$
whenever its size can be inferred from context and use $I_n$ to denote an $n \times n$ identity matrix.

Let $\eta^1_t,\ldots,\eta^n_t$ be $n$ independent vacuum quantum white noise
processes satisfying the commutation relations:
$[\eta^j_t,(\eta^k_{t'})^*]=\delta_{jk}\delta(t-t')$ and
$[\eta^j_t,\eta^k_{t'}]=0$ $\forall j,k$ and $\forall t,t' \geq
0$, and define $A^j_t=\int_{0}^t \eta^j_sds$ ($j=1,\ldots,n$) to
be a vacuum annihilation field, $A^{j*}_t=\int_{0}^t (\eta^j_s)^*ds$ to be a
vacuum creation field,  and $\Lambda_t^{jk}=\int_{0}^t (\eta^j_s)^*\eta^k_s ds$. The processes
$A^j_t,A^{j*}_t,\Lambda^{jk}_t$ are quantum stochastic processes 
collectively referred to as {\em fundamental processes}, and $A^{j*}_t = (A^j_t)^*$. 
The products of their differential increments are given by the 
quantum It\^{o} table \cite{HP84,KRP92}%
\[%
\begin{tabular}
[c]{c|ccc}%
$dM_{t}^{1}\,\backslash\,dM_{t}^{2}$ & $  dA_{t}^{j*}$ &
$d\Lambda_{t}^{jk}$ & $dA_{t}^{j}$\\\hline
$ dA_{t}^{l*} $ & $0$ & $0$ & $0$\\
$d\Lambda_{t}^{lm}$ & $\delta_{mj} dA_{t}^{l*}  $ &
$\delta_{mj}\,d\Lambda_{t}^{lk}$ & $0$\\
$dA_{t}^{l}$ & $\delta_{lj}\,dt$ & $\delta_{lj}\,dA_{t}^{k}$ & $0$%
\end{tabular}.
\]
All products $dM_{t}\,dt$ and $dt\,dM_{t}$ vanish. If two processes $X_t$ and $Y_t$ have the quantum stochastic differentials:
\begin{eqnarray}
dX_t = \sum_{j=1}^{n}\left(F_{1j} dA_t^j + G_{1j} dA_t^j + \sum_{k=1}^n H_{1jk} d\Lambda_t^{jk}\right), \nonumber\\ 
dY_t = \sum_{j=1}^{n}\left(F_{2j} dA_t^j + G_{2j} dA_t^j + \sum_{k=1}^n H_{2jk} d\Lambda_t^{jk} \right),
\end{eqnarray}
for constant system operators $F_{lj},G_{lj},H_{ljk}$, $l=1,2$, then the quantum stochastic differential of the product $X_tY_t$ is given by the {\em quantum It\^{o} rule}:
\begin{align}
dX_tY_t = dX_t Y_t + X_t dY_t  + dX_t dY_t,
\end{align}
in which each term on the right hand side can be evaluated using the quantum It\^{o} table given above, and using the fact that $X_t$ and $Y_t$ commute with each of the differentials $dA_t^j$, $dA_t^{j*}$ and $d\Lambda_t^{jk}$ \cite{HP84,KRP92}.

Let $A_t=(A_t^1,A_t^2,\ldots,A_t^n)^T$. In the interaction picture with respect to the free field dynamics, the general evolution of an open Markov quantum system coupled to the fundamental processes at any time $t \geq 0$ is given by a unitary (co-cycle) propagator $V(t)$ satisfying the {\em right} Hudson-Parthasarathy (HP) quantum stochastic differential equation (QSDE) \cite{HP84,KRP92,Fagn90}:
\begin{eqnarray}
dV_t  &=& \left(\sum_{j,k=1}^{m}(S_{r,jk}-\delta_{jk})d\Lambda^{jk}_t
+dA_t^{\dag}L_r- L_r^{\dag}SdA_t-(iH_r +
\frac{1}{2}L_r^{\dag}L_r  )dt\right) V_t; \quad V(0)=I,\label{eq:right-QSDE} 
\end{eqnarray}
where  $H_r$ is self-adjoint (i.e., Hermitian) linear operator that is the Hamiltonian of the system, $L_r=(L_{r,1},L_{r,2},\ldots,L_{r,n})^T$ is a column vector of coupling linear operators $L_{r,j}$ to the fields, and $S_{r,jk}$, $j,k=1,\ldots,m$, is a bounded system operator such that $S_r=[S_{r,jk}]_{j,k=1,\ldots,n}$ is 
unitary, that is, $S_r^{\dag}S_r=S_rS_r^{\dag}=I$. $S_r$ is called a {\em scattering matrix}. Here we have added the subscript $r$ just to emphasize that (\ref{eq:right-QSDE}) is a right HP QSDE. If we define $K_r = -iH_r -\frac{1}{2} \sum_{k=1}^{n}  L_{r,k}^*L_{r,k}$, then we see that $K_r+K_r^*= -\sum_{k=1}^n L_{r,k}^*L_{r,k}$ and $K_r-K_r^*=-2iH_r$. Using the right propagator $V(t)$, observables evolve in the Heisenberg picture according to $a_{t}=V_{t}^*\,a_{0}V_{t}$, which has the differential form%
\begin{align}
da_{t}=V_{t}^*\,a_{0}dV_{t}+dV_{t}^*\,a_{0}V_{t}+dV_{t}^*%
\,a_{0}dV_{t}.
\end{align}
Note that the right QSDE is completely specified by  three parameters, $S_r$, $L_r$, and $H_r$.

In many parts of the calculations to follow in these appendices we will work  with {\em left} QSDE's involving a left unitary propagator $U_{t}$ such that observables evolve according to $a_{t}=U_{t}\,a_{0}U_{t}^{\,\ast}$.
We use this left form because we will be applying the  adiabatic elimination results 
of Bouten, van Handel and Silberfarb \cite{Bout08}, in which they adopted the left QSDEs in their work for
technical reasons, rather than physical ones; see \cite[Remark 3]{Bout08}. 

The relation between the left propagator $U_t$ and right propagator $V_t$ for 
the same system is
\begin{align}
dV_{t}=dU_{t}^{\ast}.
\end{align}

\subsection{The master equation}
\label{sec:master-equation} Working with the right QSDE we follow Gardiner's method  (see, e.g., \cite{GZ00}) for deriving the master equation;  see also the derivation in \cite{HP84,KRP92}. By averaging over the noise terms in the evolution, we obtain the expected evolution of an observable%
\begin{align}
d\left\langle a_{t}\right\rangle  & =\left\langle da_{t}\right\rangle
=\left\langle V_{t}^{\,\ast}\,a_{0}dV_{t}+dV_{t}\,^{\,\ast}a_{0}V_{t}%
+dV_{t}^{\,\ast}\,a_{0}dV_{t}\right\rangle ,\nonumber\\
\frac{d}{dt}\left\langle a_{t}\right\rangle  & =\mathsf{Tr}\,\left[  \left\{
V_{t}^{\,\ast}a_{0}K_rV_{t}+V_{t}^{\,\ast}K_r^{\,\ast}a_{0}V_{t}+\sum_{j=1}^{n}%
V_{t}^{\,\ast}L_{r,j}^{\,\ast}a_{0}L_{r,j}V_{t}\right\}  \rho_{0}\right].
\end{align}
This may be compared with%
\begin{align}
\frac{d}{dt}\left\langle a\right\rangle =\mathsf{Tr}\,\left[  a_{0}\frac
{d\rho_{t}}{dt}\right]  ,\quad\rho_{t}=V_{t}\rho_{0}V_{t}^{\,\ast},
\end{align}
to obtain%
\begin{align}
\frac{d\rho_{t}}{dt}  & =K_r\rho_{t}+\rho_{t}\,K_r^{\,\ast}+\sum_{j=1}^{n}%
L_{r,j}\rho_{t}L_{r,j}^{\,\ast}\nonumber\\
& =-i\left[  H_r,\rho_{t}\right]  +\sum_{j=1}^n\left\{  L_{r,j}\rho_{t}%
L_{r,j}^{\,\ast}-\frac{1}{2}L_{r,j}^{\,\ast}L_{r,j}\rho_{t}-\frac{1}{2}\rho_{t}L_{r,j}^{\,\ast}L_{r,j}\right\}  .
\end{align}

\subsection{Concatenation and series product}
\label{sec:concat-series-prods}
Let $G$  denote an open Markov quantum system  that evolves according to
the right QSDE (\ref{eq:right-QSDE}) with given parameters $S_r$, $L_r$ and
$H_r$. For compactness, we shall use a shorthand notation
of \cite{Goug09a}, and denote such a system as
$G=(S_r,L_r,H_r)$. In this section we briefly recall the
concatenation and series product developed in \cite{Goug09a} that
allows one to systematically obtain the parameters of an
open Markov quantum system built up from  certain interconnections of
open Markov quantum systems.

Let $G_1=(S_1,L_1,H_1)$ and
$G_2=(S_2,L_2,H_2)$ be two open Markov quantum systems. 
The concatenation product $G_1 \boxplus G_2$ of $G_1$ and $G_2$ is defined as
\begin{equation}
G_1\boxplus G_2=(\left[\begin{array}{cc} S_1 & 0 \\ 0 &
S_2\end{array}\right],\left[\begin{array}{c} L_1\\L_2 \end{array} \right],H_1+H_2).
\end{equation}
It is important to note here that elements of $S_1,L_1,H_1$
need not commute with those of $S_2,L_2,H_2$ (i.e., $G_1$ and $G_2$ may be the sub-systems of a common larger system). If $G_1$ and $G_2$ are independent systems,
in the sense that their coefficients commute with one another then  concatenation can be interpreted
simply as the grouping of variables of two
non-interacting open Markov systems to form a larger
open Markov system.

It is also possible to feed the output of a system $G_1$ (here the output is the field that results
after the incoming field has interacted with the system; see \cite{GZ00,Goug09a} for details) to the
input of system $G_2$, with the proviso that $G_1$ and $G_2$ have
the same number of input and output channels. This operation of
cascading or loading of $G_2$ onto $G_1$ is represented by the
series product $G_2 \triangleleft G_1$ defined by:
\begin{eqnarray}
G_2 \triangleleft G_1 &=&(S_2S_1,S_2 L_1+L_2,H_1+H_2+\Imag\{L_2^{\dag}S_2 L_1\}).
\end{eqnarray}
Note that $G_2 \triangleleft G_1$ is again an open Markov quantum system with a scattering matrix, coupling operator and
Hamiltonian as given by the above formula. Moreover, there are two important decompositions of an open Markov quantum system based on the series product. They are
\begin{eqnarray}
(S,L,H)&=&(I,L,H) \triangleleft (S,0,0)=(S,0,0) \triangleleft (I,S^{\dag}L,H).
\end{eqnarray} 

\subsection{Coherent fields and the Weyl operator}
\label{sec:Weyl-operators}
Thus far, we have provided overviews for QSDEs driven by quantum noise fields that are in the vacuum state. However, in many situations in quantum optics the fields are in coherent states, emanating from a laser source.  Suppose that each of the fields $A_t^1,A_t^2,\ldots,A_t^n$ is in a coherent state with (complex) amplitude $\alpha_1,\alpha_2,\ldots,\alpha_n$, and let $\alpha=(\alpha_1,\alpha_2,\ldots,\alpha_n)^T$; note that this includes the vacuum states by setting the amplitude for a particular field to zero. Fields with coherent states can be obtained from fields in the vacuum state by an application of a Weyl operator $W^{\alpha}_t$ satisfying {\em both} the right and left QSDE:
\begin{align}
dW^{\alpha}_t  &= (dA_t^{\dag}\alpha - \alpha^{\dag} dA_t -\frac{1}{2}|\alpha|^2dt)W_t^{\alpha},\nonumber\\
			&= W_t^{\alpha} (dA_t^{\dag}\alpha - \alpha^{\dag} dA_t -\frac{1}{2}|\alpha|^2dt),
\end{align}
and we also have the relation
\begin{align}
W_t^{\alpha *} = W_t^{-\alpha}.
\end{align}

Then the right unitary propagator $V_t^{\alpha}$ of a system driven by fields with coherent amplitudes $\alpha$ is $V_t^{\alpha}=V_t W^{\alpha}_t$, where $V_t$ is the right QSDE given in (\ref{eq:right-QSDE}), representing the same system with vacuum inputs.  Alternatively, using the series product formalism from the preceding subsection, we can write $G_{\alpha}=G \triangleleft (I,\alpha,0)$, where $G_{\alpha}=(S,L_r+S\alpha, H_r + \Imag\{L^{\dag}S\alpha\})$ is the system driven by coherent fields with amplitudes $\alpha$, and $G$ is the system when driven by vacuum fields. 

\section{Component and subsystem models}
\label{sec:subsystems-derivations}
In this section we describe several of the network subsystems using the formalism introduced in section \ref{sec:prelim}.  We also introduce and apply a recent adiabatic elimination theorem \cite{Bout08} that generates a QSDE description of an often much simpler, approximate evolution of a physical system with well-separated time scales.  These approximate models form the subsystem components of the full network model presented in section \ref{sec:network}.
 
\subsection{Probe component model}
\label{sec:probe-derivation}

In this section we present the physical cQED model for the probed qubits and derive its limiting evolution used in the network model \cite{Kerc09a}.  

As represented by the blue transitions in Fig. \ref{fig:levels}, the $Z$ and $X$ probe interactions emerge from a three level atomic system in a $\lambda$-configuration interacting with a quantized mode of a single-sided electromagnetic resonator (cavity).  In the case of the $Z$ probe, the $\vert g\rangle\leftrightarrow\vert e\rangle$ transition is strongly coupled and on resonance with the cavity mode, while the $\vert h\rangle$ state is fully uncoupled.  In the case of the $X$ probe, both $\vert g\rangle\leftrightarrow\vert e\rangle$ and $\vert h\rangle\leftrightarrow\vert e\rangle$ transitions are simultaneously coupled and on resonance with the cavity mode.  The $X$ probe configuration could be realized with a degenerate spin-$\frac12$ ground state pair coupled to an excited spin-0 state via a linearly polarized cavity mode, for example.  Thus, in a properly rotating frame and applying the usual rotating wave approximation, the familiar Jaynes-Cummings Hamiltonian interaction for both systems is 
\begin{equation}
H = ig_c(\sigma ^* a-\sigma a ^*)
\end{equation} 
where $a$ is the cavity mode annihilation operator, the atomic `lowering operators' act as $\sigma= \vert g\rangle\langle e\vert$ on the atomic states in the bit-flip network, $\sigma= \frac{1}{\sqrt2}\left(\vert g\rangle+\vert h\rangle\right)\langle e\vert$ in the phase-flip network, and the coupling rate $g_c$ may be taken real without loss of generality.

However, the cavity mode and atom also interact with the free fields that surround them: the cavity photons will decay out of the resonator and the excited atomic state will spontaneously emit photons into the environment.  These interactions are included in a HP QSDE description of the dynamics with the inclusion of two coupling operators \footnote{The physical system suggested for the X probe should actually include three coupling operators, as the state $\vert e\rangle$ may spontaneously emit right- or left-handed photons in a given direction.  However, the proper inclusion of this additional decay channel leads to an equivalent limiting evolution.}:
\begin{equation}
L_1 = \sqrt{2\kappa}a;\,\,L_2 = \sqrt{2\gamma_\perp}\sigma
\end{equation}
where $\kappa$ and $\gamma_\perp$ are the cavity field and atomic dipole decay rates, respectively.  As there is no direct coupling between the cavity and atomic decay channels, the scattering matrix for this system is the identity.  Thus, the right QSDE descriptions of the dynamics for the $Z$ and $X$ systems \emph{with vacuum inputs} are characterized by the operator coefficients
\begin{equation}
Q = \left(I,\left[\begin{tabular}{c}$\sqrt{2\kappa}a$\\ $ \sqrt{2\gamma_\perp}\sigma$\end{tabular}\right], ig_c(\sigma ^* a-\sigma a ^*)\right).
\end{equation}

We will eventually use the limiting evolution of these `vacuum' cQED systems in our network model, but note that familiar, \emph{driven} cQED dynamics may be derived with the inclusion of a Weyl operator in series product with the vacuum system (representing an on-resonance, CW coherent-state input to the cavity mode)
\begin{eqnarray}
Q^{\alpha} &=&  Q\triangleleft\left(I,\left[\begin{tabular}{c}$\alpha$\\ $ 0$\end{tabular}\right], 0\right) \nonumber\\
&=&  \left(I,\left[\begin{tabular}{c}$\sqrt{2\kappa}a+\alpha$\\ $ \sqrt{2\gamma_\perp}\sigma$\end{tabular}\right], ig_c(\sigma ^* a-\sigma a ^*)+\frac{i\sqrt{\kappa}}{2}(\bar\alpha a - \alpha a ^*)\right),
\end{eqnarray}
followed by the application of the master equation from section \ref{sec:master-equation}.

We now begin an adiabatic elimination procedure on $Q$ to obtain the very simple limiting dynamics of a measurable qubit, which will be used in the network model.  The physical intuition of the derived dynamics is as follows.  As described in \cite{Kerc09a}, in a bad cavity limit ($\kappa\rightarrow\infty$), with the atom in a uncoupled state, a coherent probe resonant with the cavity will enter the cavity and exit `immediately' (due to the large $\kappa$), picking up no additional phase shift.  However, if the atom is in a coupled state, and the system is also in a `small volume' coupling limit ($g_c,\kappa\rightarrow\infty$, with $g_c/\kappa$
constant), then the same probe will simply reflect from the off-resonant, atom-cavity system, picking up an additional $\pi$ phase shift.  Thus, in these limits, the cQED system acts as a simple scattering object with no internal dynamics, but which imparts a phase shift on a cavity-incident coherent probe, as determined by an internal qubit state.

Following the mathematical formalism of \cite{Bout08}, we redefine and relabel $Q$ with a scaling parameter $k$, which is assumed to take the limit $k\rightarrow\infty$, and write the component as a left QSDE
\begin{equation}\begin{split}
dQ_{t}^{(k)} = Q_{t}^{(k)}\Big\{-k\sqrt{2\kappa}adA_t^{1*}+k\sqrt{2\kappa}a^*dA_t^{1}-\sqrt{2\gamma_\perp}\sigma dA_t^{2*}+\sqrt{2\gamma_\perp}\sigma^*dA_t^{2}-\nonumber\\
k^2\kappa a^*adt-\gamma_\perp\sigma^*\sigma dt-k^2g_c(\sigma^* a-\sigma a ^*)dt\Big\}.
\end{split}\end{equation}      
Note that with the scaling, the cavity field decay and coupling rates take the limits described above.

 In the notation of \cite{Bout08}, the operator coefficients of the above left QSDE are%
\begin{align}
K^{(k)}=-k^{2}\kappa a^{\ast}a-\gamma_\perp\sigma^{\ast}\sigma%
-k^{2}g_c\left(  a^{\ast}\sigma -a\sigma^{\ast}\right)  ,\quad
L^{(k)}=\left(
\begin{array}
[c]{l}%
k\sqrt{2\kappa}a^{\ast}\\
\sqrt{2\gamma_\perp}\sigma^{\ast}%
\end{array}
\right)  ,\quad N^{(k)}=I.
\end{align}
Through the adiabatic elimination procedure of \cite{Bout08}, a set of limiting operator coefficients (which we denote without the `$^{(k)}$') are derived, which define another, limiting propagator, $Q_{t}$, to which $Q_{t}^{(k)}$ strongly converges.  It is this limiting propagator that will be used in the network model in place of the physical cQED model.

For the adiabatic elimination \cite[section 2.2]{Bout08} we use 
\begin{align}
Y  & =-\kappa a^{\ast}a-g_c\left(  a^{\ast}\sigma-a\sigma^{\ast}\right)  ,\quad
A=0,\quad B=-\gamma_\perp\sigma^{\ast}\sigma,\nonumber\\
F  & =\left(
\begin{array}
[c]{l}%
\sqrt{2\kappa}a^{\ast}\\
0
\end{array}
\right)  ,\quad G=\left(
\begin{array}
[c]{l}%
0\\
\sqrt{2\gamma_\perp}\sigma^{\ast}%
\end{array}
\right)  ,\quad W=I.
\end{align}
In the $Z$ probe case for the bit-flip network, we choose $H_{0}=\mathsf{span}\left\{  \left\vert \,g,0\right\rangle
,\left\vert \,h,0\right\rangle \right\}  $ and define%
\begin{align}
\tilde{Y}\,\left\{  \left\vert \,g\,0\right\rangle ,\left\vert
\,h\,0\right\rangle \right\}  &=0,\nonumber\\
\tilde{Y}\,\left\vert \,h\,n\right\rangle  & =-\frac{1}{\kappa n}\left\vert
\,h\,n\right\rangle ,\quad n\geq1;\nonumber\\
\tilde{Y}\,\left\vert \,e\,\left(  n-1\right)  \right\rangle  & =-\frac
{g_c\sqrt{n}}{\kappa^{2}n\left(  n-1\right)  +g_c^{2}n}\left\vert
\,g\,n\right\rangle -\frac{\kappa n}{\kappa^{2}n\left(  n-1\right)  +g_c^{2}%
n}\left\vert \,e\,\left(  n-1\right)  \right\rangle ,\quad n\geq1;\nonumber\\
\tilde{Y}\,\left\vert \,g\,n\right\rangle  & =-\frac{\kappa\left(  n-1\right)
}{\kappa^{2}n\left(  n-1\right)  +g_c^{2}n}\left\vert \,g\,n\right\rangle
+\frac{g_c\sqrt{n}}{\kappa^{2}n\left(  n-1\right)  +g_c^{2}n}\left\vert
\,e\,\left(  n-1\right)  \right\rangle ,\quad n\geq1.
\end{align}
This provides $Y\tilde{Y}=\tilde{Y}Y=P_{1}$. Then by \cite{Bout08}, the limiting operator coefficients are,%
\begin{align}
 K =0,\quad  L_{1}= L_{2}=0,\quad  M_{1}= M_{2}=0,
\end{align}
and (with $\Pi_{in} = \vert i,n\rangle\langle i,n\vert$)%
\begin{align*}
 N_{11}  & =P_{0}W_{11}\left(  F_{1}^{\ast}\tilde{Y}F_{1}+\delta
_{11}\right)  P_{0}=P_0+2\kappa P_{0}a\tilde{Y}a^{\ast}\left(  \Pi_{g0}+\Pi
_{h0}\right) \\
& =P_0+2\kappa P_{0}a\tilde{Y}\left(  \left\vert \,g\,1\right\rangle
\left\langle g\,0\right\vert +\left\vert \,h1\right\rangle \left\langle
h\,0\right\vert \right) \\
& =P_0+2\kappa P_{0}a\left(  \frac{1}{g_c}\left\vert \,e\,0\right\rangle\left\langle g\,0\right\vert -\frac{1}{\kappa}\left\vert \,h\,1\right\rangle
\left\langle h\,0\right\vert \right) \\
& =P_0-2\left\vert \,h\,0\right\rangle \left\langle h\,0\right\vert \\
& =\Pi_{g0}-\Pi_{h0} \equiv Z,
\end{align*}%
\[
 N_{12}=P_{0}W_{11}\left(  F_{1}^{\ast}\tilde{Y}F_{2}+\delta
_{12}\right)  P_{0}=0,
\]%
\[
 N_{21}=P_{0}W_{22}\left(  F_{2}^{\ast}\tilde{Y}F_{1}+\delta
_{21}\right)  P_{0}=0,
\]%
\begin{align}
N_{22}=P_{0}W_{22}\left(  F_{2}^{\ast}\tilde{Y}F_{2}+\delta
_{22}\right)  P_{0}=1,
\end{align}
Collecting the results, the limiting system is characterized by the operator coefficients%
\begin{equation}\label{eq:qubitlimit}
Q = \left(\left[\begin{tabular}{c c}
Z & 0\\
0 &I\end{tabular}\right],\left[\begin{tabular}{c}$0$\\ $ 0$\end{tabular}\right], 0\right),
\end{equation}
to which the physical cQED system, $Q^{(k)}$, strongly converges in the limit $k\rightarrow\infty$.  For the $X$ probe case in the phase-flip network, the analogous follows, but with the pauli-$X$ spin operator eventually replacing $Z$ in equation \eqref{eq:qubitlimit}. 

\subsection{Raman transition model}
\label{sec:Raman-derivation}

In this section we describe the physical Raman interaction that is used to correct the qubit states in the coherent feedback loop.  We presume in this case that the atoms in the probed qubits described in subsection \ref{sec:probe-derivation} are also interrogated by two free field modes (i.e. field modes transverse to the cavity mode) that drive atomic Raman transitions when displaced by a large, coherent amplitude.

We assume that there also exists an additional atomic excited state $\vert r\rangle$, distinct and far detuned from $\vert e\rangle$ for theoretical simplicity, although $\vert r\rangle=\vert e\rangle$ should work in practice.   Both ground states are coupled to $\vert r\rangle$ via interactions with transverse free fields.  Two of these field modes, far detuned by frequency  $\Delta$ from the $\{\vert h\rangle,\vert g\rangle\}\leftrightarrow\vert r\rangle$ transition are used to drive the Raman interaction, as represented by the red transitions in Fig. \ref{fig:levels}.  For the bit-flip network, in a properly rotating frame, these interactions are modeled in a QSDE by the Hamiltonian and coupling operators (redefining $\Delta\rightarrow k^2\Delta$ in anticipation of an eventual $k\rightarrow\infty$ limit)\footnote{All atomic operators in this section implicitly act as the identity on the cavity mode.}
\begin{eqnarray}
H^{(k)} &=& k^2\Delta\vert r\rangle\langle r\vert  \equiv k^2\Delta\Pi_r\nonumber\\ 
L_1^{(k)} &=& \sqrt{\gamma}\vert h\rangle\langle r\vert \equiv \sqrt{\gamma}\sigma_{hr}\nonumber\\
L_2^{(k)} &=& \sqrt{\gamma}\vert g\rangle\langle r\vert \equiv \sqrt{\gamma}\sigma_{gr}.
\end{eqnarray}
For the phase-flip network, the above coupling operators are modified with $\sigma_{hr}\rightarrow\frac{1}{\sqrt2}(\sigma_{hr}+\sigma_{gr})$ and $\sigma_{gr}\rightarrow\frac{1}{\sqrt2}(\sigma_{hr}-\sigma_{gr})$ (physically suggesting two linearly polarized modes coupling to angular momentum eigenstates, for example).  Also, the interactions of the $\{\vert h\rangle,\vert g\rangle\}\leftrightarrow\vert r\rangle$ transitions with all other free modes (which principally drive spontaneous emission decay from $\vert r\rangle$ to the ground states) may be aggregated into just two more coupling operators
\begin{eqnarray}
L_3^{(k)} &=& \sqrt{\gamma_\parallel}\sigma_{hr}\nonumber\\
L_4^{(k)} &=&  \sqrt{\gamma_\parallel}\sigma_{gr}
\end{eqnarray}
in the limit $\gamma\ll\gamma_\parallel$.  Together with an identity scattering matrix, the above Hamiltonian and coupling operators represent a simple vacuum-input Raman system, $Q_{R}^{(k)}$.  

In the quantum network, the field modes 1 and 2 undergo a non-trivial processing before reaching the Raman systems described above.  For demonstration purposes, however, consider a bit-flip Raman system in isolation, but with modes 1 and 2 simply displaced into large-amplitude coherent states.  This system can be represented by the operator coefficients
\begin{eqnarray}
Q_R^{\beta,(k)} &=& Q_{R}^{(k)}\triangleleft \left(W_{k\beta_1}\boxplus W_{k\beta_2} \boxplus (I_2,0,0)\right)\nonumber\\
&=& \left(I,\left[\begin{tabular}{c}
$\sqrt{\gamma}\sigma_{hr}+k\beta_1$\\ 
$ \sqrt{\gamma}\sigma_{gr}+k\beta_2$\\
$\sqrt{\gamma_\parallel}\sigma_{hr}$\\
$\sqrt{\gamma_\parallel}\sigma_{gr}$
\end{tabular}\right], k^2\Delta\Pi_r+\frac{ik\sqrt{\gamma}}{2}(\sigma_{hr}\beta_1^*-\sigma_{hr}^*\beta_1)+\frac{ik\sqrt{\gamma}}{2}(\sigma_{gr}\beta_2^*-\sigma_{gr}^*\beta_2)\right)
\end{eqnarray}
where $W_{k\beta}$ is a Weyl operator of displacement amplitude $k\beta$.

Application of the adiabatic elimination theorem \cite{Bout08} to the above (using a technique discussed below in section \ref{sec:Luc's-trick}) results in an approximate Hamiltonian interaction 
\begin{equation}\label{eq:RamanH}
H = -\frac{\gamma}{\Delta}\left(\vert \beta_1\vert^2\Pi_h+ \vert \beta_2\vert^2\Pi_g + \beta_1\beta_2^*\sigma_{gh}+\beta_1^*\beta_2\sigma_{gh}^*\right)
\end{equation}
and all atomic operators drop out of the field coupling operators (eliminating all spontaneous emission dynamics).  Thus, coherent transitions between the two ground states may be driven by the last two terms in H when both Raman modes are in large-amplitude coherent states.  However, the ground state energy shifts represented by the first two terms (physically recognizable as AC Stark shifts) are generally inconvenient in the QEC network.  

To compensate for these Stark shifts, in principle we could presume an additional pair of excited states $\{\vert H\rangle,\vert G\rangle\}$ (not shown in Fig. \ref{fig:levels}) that also couple to the ground states $\{\vert h\rangle,\vert g\rangle\}$ via modes 1 and 2, respectively, but with the opposite detunings: $-k\Delta$.  This choice to add more atomic levels is again made for theoretical simplicity, although two additional free field modes coupling to $\vert r\rangle$, but with opposite detunings, would likely be more practical.  Thus, the physical Raman interaction would have the form
\begin{equation}
Q_{R^+}^{(k)} = \left(I_2,\left[\begin{tabular}{c}
$\sqrt{2\gamma}(\sigma_{hr}+\sigma_{Hr})$\\ 
$ \sqrt{2\gamma}(\sigma_{gr}+\sigma_{gG})$\\
\end{tabular}\right], k^2\Delta(\Pi_r-\Pi_H-\Pi_G)\right)
\end{equation}
where we have omitted the coupling of the excited states to any other free modes, as the dynamics  they induce would eventually drop out in the $k\rightarrow\infty$ limit, as above.  In this limit, the effective evolution is again governed by the Hamiltonian dynamics depicted in equation \eqref{eq:RamanH}, but without the first two, AC Stark shift terms.

\subsection{Adiabatic elimination with large field displacements}
\label{sec:Luc's-trick}

In the previous section we mentioned that the adiabatic elimination processes will be applied to the Raman components only after they are placed in the quantum network.  However, there are some mathematical complications in applying the usual theorem \cite{Bout08} when the amplitude of coherent field input must be scaled.  In this section we describe a procedure to overcome this difficulty briefly and in the abstract.  This technique is a generalization upon a similar technique developed by Luc Bouten and applied in \cite{Kerc09a}.

Let $(S^0,L^0,H^0)$ be a network with only vacuum inputs. The same network, but with coherent displaced inputs is defined as 
\begin{eqnarray}
G^{(k)} &=& (S^0,L^0,H^0)\triangleleft(I,d_k,0) \nonumber\\
&=&(S^0,S^0d_k+L^0,H^0+\Im\{L^{0\dag} S^0d_k\})= (S,L^{(k)},H^{(k)})
\end{eqnarray}
where $d_k$ is a vector with complex-valued entries, some of which may scale with $k$ (e.g. $k\beta$), some may not (e.g. $2\alpha$), and some still may be 0.  The scattering matrix $S=S^0$ is assumed to have no scaling with $k$.  It happens that applying the adiabatic elimination theorem on $G^{(k)}$ is complicated by the $S^0d_k$ term in the coupling operators.  The first step in overcoming these difficulties is to reconstruct the network such that all the scaled parameters are moved to the Hamiltonian of a new subcomponent.  That is, we rewrite
\begin{eqnarray}\label{eq:Luc's-trick}
G^{(k)} &=& (S,0,0)\triangleleft(I,d_k,0)\triangleleft(I,S^\dag L^0,H^{(k)} - \Im\{d_k^\dag S^{0\dag} L^0\})\nonumber\\
&\equiv& (S,0,0)\triangleleft(I,d_k,0)\triangleleft \tilde{G}^{(k)}
\end{eqnarray} 
It will be straightforward to apply the adiabatic elimination procedure on the subsystem $\tilde{G}^{(k)}$, which will produce a limiting subsystem $\tilde G$.  Thus $G^{(k)}$ is asympotically approximated by
\begin{equation}
\hat{G}^{(k)} =  (S,0,0)\triangleleft(I,d_k,0)\triangleleft \tilde{G}
\end{equation}
for  sufficiently large values of $k$.  It is this asymptotically approximate and simpler network $\hat G^{(k)}$ that will be used in final description of the coherent feedback system.

\section{The quantum network}\label{sec:network}
\begin{figure}[h!]
\centering
\includegraphics[scale=0.6]{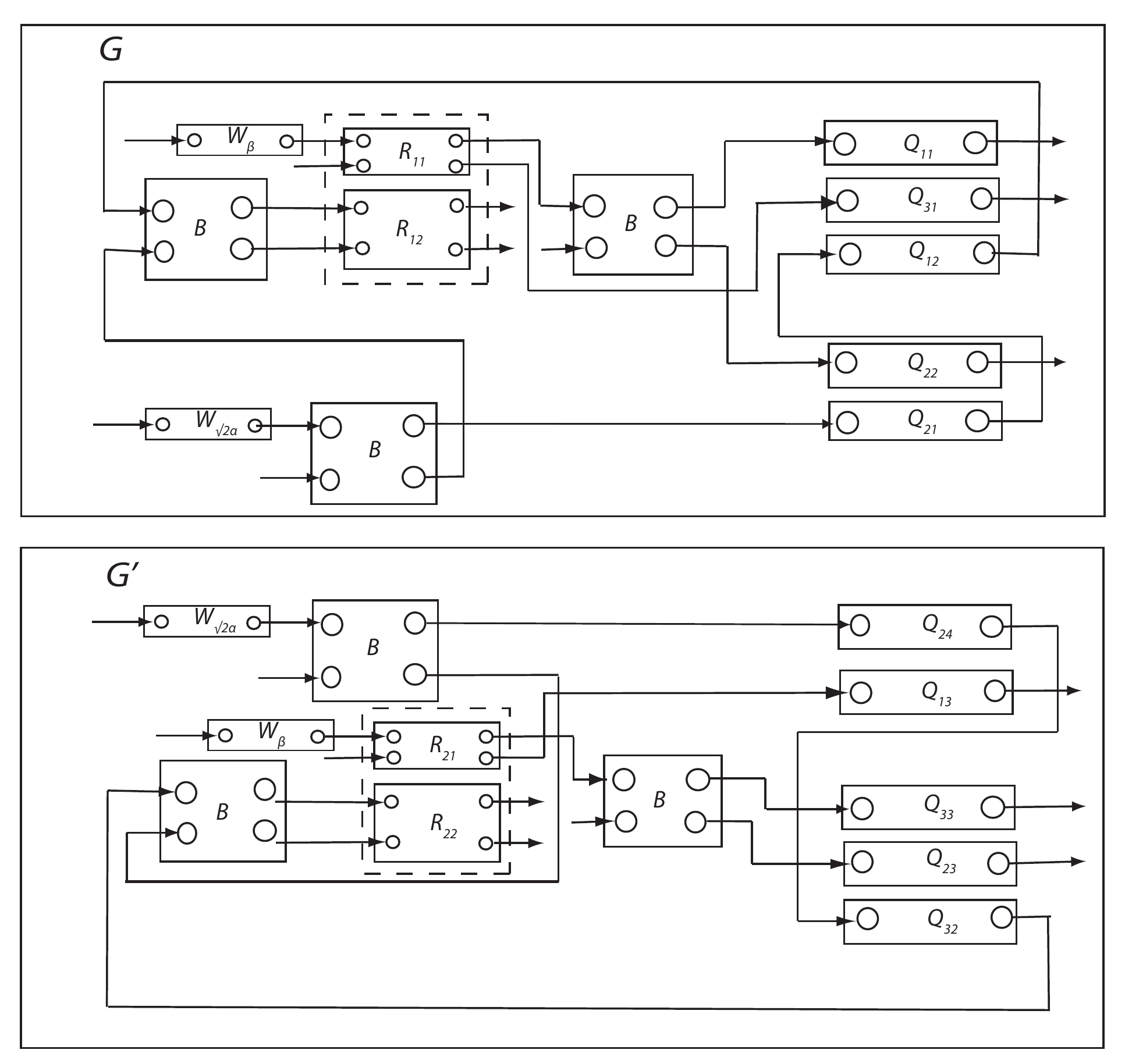}
\caption{{\bf The quantum `correction' network model}.  Concatenation decomposition of a quantum network analogous to that represented in Fig. \ref{fig:relaynet}  as $G \boxplus G'$.  See Remark \ref{rem:different-networks} for an explanation of the network differences.
} \label{fig:QEC-net-decom}
\end{figure}

Here we present a detailed calculation of the coherent-feedback network model.  A schematic diagram of the overall quantum network is shown in Fig.~\ref{fig:QEC-net-decom}.  We start by introducing notations and  reviewing the component models; 
detailed derivations for several of the component models are collected in section \ref{sec:subsystems-derivations}. 

\begin{remark}
\label{rem:different-networks}
The network considered in the appendices is slightly different from the network shown in Fig.~\ref{fig:relaynet}. This is because in these appendices, the cavity for Q2 is taken to be two sided with a coherent probe laser impinging on each side, allowing us to obtain a decomposition of the network into two (symmetric) parts as shown in Fig.~\ref{fig:QEC-net-decom} and simplifying the calculations for the overall network parameters that are presented in subsection \ref{sec:net-calculations}. On the other hand, in Fig.~\ref{fig:relaynet} we make all qubit cavities identical and one sided, which would be probably more suited in actual cQED implementations. Similar effects can be achieved with either a double sided cavity for $Q2$ or a single sided cavity that is followed by a 50/50 beamsplitter. Although the two networks are slightly different, they only differ in the overall network scattering matrix, and a slightly modified repeat of the lengthy calculations of subsection \ref{sec:net-calculations}, with $Q2$ one-sided and followed by a beamsplitter, will show that the network coupling vector and Hamiltonian will be the same for both networks. As a consequence, we emphasize that \emph{the two networks will have the same network master equation}. Since, our main interest here is the network master equation, to derive this equation it is more convenient to work with the slightly modified network of Fig.~\ref{fig:QEC-net-decom}.
\end{remark} 

\subsection{Notation}
We shall use the following symbolic convention for individual components in the network: B = beam splitter, Q = qubit cavity, R = quantum relay/switch. Distinct network components that are  of the same type are distinguished by assigning to each a unique subscript as an index, and all components in the network are open Markov quantum systems described by three parameters introduce in section \ref{sec:prelim}: a unitary scattering matrix, a coupling vector, and Hamiltonian operator. Matrices and vectors can have operator entries and an entry of a matrix $M$ in row $k$ and column $l$ is denoted as $M_{kl}$. A network component $Z_{k}$ with $Z \in \{B,C,R\}$ and $k$ some index will be written as $Z_k=(S^{(Z_k)},L^{(Z_k)},H^{(Z_k)})$, where $S^{(Z_k)},L^{(Z_k)},H^{(Z_k)}$ are, respectively, the the scattering matrix, coupling vector, and Hamiltonian operator of $Z_k$. 

If the matrix  $S^{(Z_k)}$ is diagonal of the form ${\rm diag}(S^{(Z_k)}_{11},S^{(Z_k)}_{22},\ldots,S^{(Z_k)}_{nn})$ (here ${\rm diag}(M_1,M_2,\ldots)$ denotes a block diagonal matrix with block diagonal elements $M_1,M_2,\ldots$) then $Z_k$ can be decomposed as $Z_k=Z_{k1} \boxplus Z_{k2} \boxplus \ldots \boxplus Z_{kn}$ with $Z_{kj}=(S^{(Z_{kj})},L^{(Z_{kj})},H^{(Z_{kj})})$ and $S^{(Z_{kj})}=S^{(Z_k)}_j$, $L^{(Z_{kj})}=L^{(Z_k)}_j$ and $H^{(Z_{kj})}$ satisfying  $\sum_{j=1}^{n} H^{(Z_{kj})} = H^{(Z_k)}$. We will exploit such a concatenation decomposition whenever they can be made to simplify the network calculations. Recall that it is {\em not} required that entries of the parameters of $Z_{kj}$  commute whenever $j \neq k$ and series products can be formed from such components. 

It is important to note that the series product is only defined for systems that have an equal number of input and output fields. However, naturally it is often necessary to make interconnections between systems with different numbers of input and output field fields. To apply the series product we augment the systems to be connected with ``dummy'' trivial couplings to additional fields as required to make the number of input-output fields of systems to be connected be the same. This is best illustrated with a simple example. Suppose we wish to connect a system $C_1=(S^{(C_1)},L^{(C_1)},H^{(C_1)})$  having only single input and output fields  with another system  $C_2=(S^{(C_2)},L^{(C_2)},H^{(C_2)})$ having two input-output fields by passing the output field of $C_1$ as the second field into $C_2$. Then two systems  connected in this way is given by the series product $ C_2 \triangleleft ((1,0,0) \boxplus C_1))$, where $(1,0,0)$ represents a dummy coupling of $C_1$ to a second field that merely ``passes through'' $C_1$, without interacting with it, before entering $C_2$ as the first input field. In general, all that is needed is to insert these trivial $(1,0,0)$ systems in the appropriate locations before making the series connection; see \cite{Goug09a} for several examples. 

Recall that $I_{n}$ denotes the $n \times n$ identity matrix and $I$ and $0$ to denote identity and a matrix will all zero entries, respectively, whose dimensions can be determined from context. In order to reduce the complexity of expressing trivially augmented systems we introduce the additional notation $\boxplus_k$ to denote the operation of augmenting a system {\em on the left} of $\boxplus_k$ by inserting a trivial $(I_n,0,0)$ system as the $k$-th to $(k+n-1)$-th component of the system. To illustrate,  for the system $C_{2}$ that we had before we would have:
\begin{eqnarray}
(C_2 \boxplus_2 (1,0,0)) \boxplus_4 (1,0,0) &=& \left( \left[\begin{array}{ccc} S^{(C_{2})}_{11}  & 0 & S^{(C_{2})}_{12}  \\
0 & 1 & 0\\
S^{(C_{2})}_{21} & 0 & S^{(C_{2})}_{12}
\end{array}\right] ,\left[\begin{array}{c} L^{(C_2)}_1 \\ 0 \\ L^{(C_2)}_2 \end{array} \right], H^{(C_2)} \right) \boxplus_4 (1,0,0)\nonumber\\
 &=& \left( \left[\begin{array}{cccc} S^{(C_{2})}_{11}  & 0 &  S^{(C_{2})}_{12}  & 0 \\
0 & 1 & 0 & 0\\
S^{(C_{2})}_{21} & 0 &  S^{(C_{2})}_{12} & 0\\
0 & 0 & 0 & 1 \\
\end{array}\right], \left[\begin{array}{c} L^{(C_2)}_1 \\ 0 \\ L^{(C_2)}_2 \\  0 \end{array} \right], H^{(C_2)} \right), 
\end{eqnarray}
and so on in this fashion.

\normalsize 
\subsection{Network components}

\subsubsection{Qubit cavities}
We note that in our quantum network all qubit cavities will have a concatenation decomposition. In what follows, $Q2$ will be taken to be a double ended cavity which is connected to two probe lasers and two bit correcting lasers (cf. Remark \ref{rem:different-networks}). It has a concatenation decomposition:  $Q_2=Q_{21} \boxplus Q_{22} \boxplus Q_{23} \boxplus Q_{24}$, where $Q_{21}$ and $Q_{24}$ are subsystems associated with the coupling of $Q2$ to the probe lasers, while $Q_{22}$ and $Q_{23}$  are associated with the coupling of $Q2$ to the bit correcting lasers that drive qubit Raman transitions. $Q1$ and $Q3$ are single ended cavities decomposable as $Q_{j}=Q_{j1}\boxplus Q_{j2} \boxplus Q_{j3}$ ($j=1,3$) with $Q_{j1}$ and $Q_{j3}$ associated with the coupling of $Qj$ to the two bit correcting lasers, while $Q_{j2}$ is the subsystem associated with the coupling of the cavity to a probe laser. Detailed derivations for the simplified cavity model can be found in subsections \ref{sec:probe-derivation} and \ref{sec:Raman-derivation}.

\subsubsection{Probe interactions}
Subsystems of the qubit cavities corresponding to Z-probe interactions have the following limit model for the atom-cavity
interaction: 
\begin{align}
Q_{12} =\left(  Z^{\left(  Q1\right)  },0,0\right) ,\; Q_{21} =\left(  Z^{\left(  Q2\right)  },0,0\right),\; 
Q_{24} =\left(  Z^{\left(  Q2\right)  },0,0\right),\; Q_{32} =\left(  Z^{\left(  Q3\right)  },0,0\right).
\end{align}
The analogous X-probe qubit cavities replace the substitution $Z^{(Qi)}\rightarrow X^{(Qi)}$.  Detailed derivation of the probe interaction model is provided in subsection \ref{sec:probe-derivation}.

\subsubsection{Raman transitions}\label{sec:Raman}
The bit-flip network Raman systems are:
\begin{align}
Q_{11}  & =\left(  1,\sqrt{\gamma}\sigma_{gr}^{\left(  Q1\right)
} ,\frac12\Delta\Pi
_{r}^{\left(  Q1\right)  }\right)  ,\quad
Q_{13}  =\left(  1,\sqrt{\gamma}\sigma_{hr}^{\left(  Q1\right)
} ,\frac12\Delta\Pi
_{r}^{\left(  Q1\right)  }\right) ,\nonumber\\
Q_{22}  & =\left(  1,\sqrt{\gamma} \sigma_{hr}^{\left(  Q2\right)
}  ,\frac12\Delta\Pi
_{r}^{\left(  Q2\right)}  \right)  ,\quad
Q_{23}  =\left(  1,\sqrt{\gamma}\sigma_{gr}^{\left(  Q2\right)
}  ,\frac12\Delta\Pi
_{r}^{\left(  Q2\right)  } \right)  ,\nonumber\\
Q_{31}  & =\left(  1,\sqrt{\gamma}\sigma_{gr}^{\left(  Q3\right)
} ,\frac12\Delta\Pi
_{r}^{\left(  Q3\right)  }  \right)  ,\quad
Q_{33}   =\left(  1,\sqrt{\gamma}\sigma_{hr}^{\left(  Q3\right)
}  ,\frac12\Delta\Pi
_{r}^{\left(  Q3\right)  } \right)  ,
\end{align}
where we note that the indices are connected slightly differently for $Q_{1}$
than $Q_{2}$ and $Q_{3}$, following the diagram in Fig.~\ref{fig:QEC-net-decom}. The analogous phase-flip Raman interactions replace $\sigma_{hr}\rightarrow\frac{1}{\sqrt2}(\sigma_{hr}+\sigma_{gr})$ and $\sigma_{gr}\rightarrow\frac{1}{\sqrt2}(\sigma_{hr}-\sigma_{gr})$ in the above.  Details of the Raman transition model is provided in subsection \ref{sec:Raman-derivation}.  As explained in section \ref{sec:SS}, the particular choice of coupling operators in each of the above helps to prevent Stark shift dephasing of the protected codeword \emph{without} additional, Stark shift-compensating atomic levels.

\subsubsection{Coherent displacements}
Several vacuum inputs to the network will be displaced with Weyl operators, representing constant amplitude, coherent state inputs.  For a given displacement, $d$, such inputs are represented with:
\begin{align}
W_d &=\left(I,d,0\right);.
\end{align}

\subsubsection{Beamsplitters}
All beamsplitters in the network are taken to be 50/50, mixing the inputs of two fields:%
\begin{align}
B  &=\left(  \left[
\begin{array}
[c]{ll}%
\frac{1}{\sqrt{2}} & \frac{1}{\sqrt{2}}\\
-\frac{1}{\sqrt{2}} & \frac{1}{\sqrt{2}}%
\end{array}
\right]  ,0 ,0\right) .
\end{align}

\subsubsection{Relays} 
Relays are modeled by a simplified four-state, four mode model. In the network there are two relays $R_1$ and $R_2$, each of which has the decomposition $R_k=R_{k1}\boxplus R_{k2}$. The state of the relay is controlled by the field inputs to $R_{k2}$, while the state of the relay controls the routing of the field inputs to $R_{k1}$, hence%
\begin{eqnarray}
R_{k1}&=&\left(  \left[
\begin{array}
[c]{ll}%
\Pi_{g}^{\left(  Rk\right)  } & -\Pi_{h}^{\left(  Rk\right)  }\\
-\Pi_{h}^{\left(  Rk\right)  } & \Pi_{g}^{\left(  Rk\right)  }%
\end{array}
\right] , 
0,0\right),\nonumber\\
R_{k2} &=& \left(  \left[
\begin{array}
[c]{ll}%
\Pi_{g}^{\left(  Rk\right)  } & -\sigma_{hg}^{\left(  Rk\right)  } \\
-\sigma_{gh}^{\left(  Rk\right)  } & \Pi_{h}^{\left(  Rk\right)  }
\end{array}
\right]  ,0,0\right),
\end{eqnarray}
for $k=1,2$, where the superscripts keep track of operators on different
component states. Detailed physical modelling for such quantum relays and derivation of the simplified scattering model can be found in \cite{Mabu09b}.

\subsubsection{Bit-/phase-flip errors}
The single-qubit error processes that the network seeks to correct may be described with 
\begin{align}
E_X^{(Qi)} &= \left(I,\sqrt{\Gamma}X^{(Qi)},0\right)
\end{align}
for a bit-flip error process with mean rate $\Gamma$ on qubit $i$, and 
\begin{align}
E_Z^{(Qi)} &= \left(I,\sqrt{\Gamma}Z^{(Qi)},0\right)
\end{align}
for a phase-flip error process on qubit $i$, also with mean rate $\Gamma$.

\subsection{Network calculations}
\label{sec:net-calculations}
The main task is to calculate the `correction' network $G\boxplus G^{\prime},$ with the
subnets $G$ and $G^{\prime}$ defined as follows.

Parts of $G:$%
\begin{align}
G_{p}  & =R_{12}\vartriangleleft B\vartriangleleft\left(  \left(
Q_{12}\vartriangleleft Q_{21}\right)  \boxplus\left(  I,0,0\right)  \right)
\vartriangleleft B\triangleleft\left(W_{\sqrt{2}\alpha}\boxplus(I,0,0)\right),\nonumber\\
G_{f}  & =\left(  Q_{11}\boxplus Q_{31}\boxplus Q_{22}\right)
\vartriangleleft\left(  B\boxplus_{2}\left(  I,0,0\right)  \right)
\vartriangleleft\left(  R_{11}\boxplus\left(  I,0,0\right)  \right)\triangleleft\left(W_\beta\boxplus(I_2,0,0)\right),
\end{align}
representing the `probe' and 'feedback' network subsystems of $G=G_p\boxplus G_f$, respectively.  Correspondences in $G^{\prime}:$%
\begin{align}
R_{22}  \leftrightarrow R_{12},\;
R_{21}  \leftrightarrow R_{11},\;
Q_{23}  \leftrightarrow Q_{11},\;
Q_{33}  \leftrightarrow Q_{22},\;
Q_{13}  \leftrightarrow Q_{31},\;
Q_{24}  \leftrightarrow Q_{21},\;
Q_{32}  \leftrightarrow Q_{12}.
\end{align}
The overall structure of the network is (with a trivial rearrangement of field indices)%
\begin{eqnarray}
N&=&\left(  
\diag(S^{\left(  p\right)  },
S^{\prime\left(  p\right)  } , S^{\left(  f\right)  },  S^{\prime\left(  f\right)  }),
~\left[
\begin{array}
[c]{l}%
L^{\left(  p\right)  }\\
L^{\prime\left(  p\right)  }\\
L^{\left(  f\right)  }\\
L^{\prime\left(  f\right)  }%
\end{array}
\right] ,~H^{\left(  p\right)  }+H^{\prime\left(  p\right)  }+H^{\left(
f\right)  }+H^{\prime\left(  f\right)  }\right).
\end{eqnarray}
In this calculation, the network components representing the bit- or phase-flips errors, $E^{(Qi)}_{X,Z}$, may be simply concatenated to the `correction' network $N$ immediately before the derivation of the full system's master equation.

\subsubsection{Probe network}\label{sec:probe_network}
Plugging in component models into $G_p$ and $G_p^{\prime}$ we arrive at%
\begin{align}
G_p &=  \left(\frac12\left[\begin{array}
[c]{lll}%
O^{(12)}\Pi_g^{(R1)}+E^{(12)}\sigma^{(R1)}_{hg} & E^{(12)}\Pi_g^{(R1)}+O^{(12)}\sigma^{(R1)}_{hg}\\
-E^{(12)}\Pi_h^{(R1)}-O^{(12)}\sigma^{(R1)}_{gh} & -O^{(12)}\Pi_h^{(R1)}-E^{(12)}\sigma^{(R1)}_{gh}%
\end{array}
\right],
\frac{\alpha}{\sqrt2}\left[\begin{array}
[c]{l}%
\Pi_g^{(R1)}O^{(12)}+\sigma_{hg}^{(R1)}E^{(12)}\\
-\sigma_{gh}^{(R1)}O^{(12)}-\Pi_h^{(R1)}E^{(12)}
\end{array}\right],0\right),\nonumber\\
G_p^{\prime} &=  \left(\frac12\left[\begin{array}
[c]{lll}%
O^{(32)}\Pi_g^{(R2)}+E^{(32)}\sigma^{(R2)}_{hg} & E^{(32)}\Pi_g^{(R2)}+O^{(32)}\sigma^{(R2)}_{hg}\\
-E^{(32)}\Pi_h^{(R2)}-O^{(32)}\sigma^{(R2)}_{gh} & -O^{(32)}\Pi_h^{(R2)}-E^{(32)}\sigma^{(R2)}_{gh}%
\end{array}
\right],
\frac{\alpha}{\sqrt2}\left[\begin{array}
[c]{l}%
\Pi_g^{(R2)}O^{(32)}+\sigma_{hg}^{(R2)}E^{(32)}\\
-\sigma_{gh}^{(R2)}O^{(32)}-\Pi_h^{(R2)}E^{(32)}
\end{array}\right],0\right),
\end{align}
where, for the bit-flip correcting network, $E^{(ij)} = Z^{(Qi)}Z^{(Qj)}+1$ and $O^{(ij)} = Z^{(Qi)}Z^{(Qj)}-1$ are proportional to two-qubit parity projectors.  For the phase flip network, $E^{(ij)} = X^{(Qi)}X^{(Qj)}+1$ and $O^{(ij)} = X^{(Qi)}X^{(Qj)}-1$.

\subsubsection{Feedback network}\label{sec:feedback_network}
Plugging in the bit-flip component models into $G_f$ and $G_f^{\prime}$ we arrive at
\begin{align}
G_f &= \left(\frac{1}{\sqrt2}\left[\begin{array}
[c]{lll}%
\Pi_g^{(R1)} & -\Pi_h^{(R1)} & 1\\
-\sqrt{2}\Pi_h^{(R1)} & \sqrt{2}\Pi_g^{(R1)} & 0\\
-\Pi_g^{(R1)} & \Pi_h^{(R1)} & 1%
\end{array}\right],\left[\begin{array}
[c]{l}%
\sqrt{\gamma}\sigma_{gr}^{(Q1)}+\frac{1}{\sqrt2}\beta\Pi_g^{(R1)}\\
\sqrt{\gamma}\sigma_{gr}^{(Q3)}-\beta\Pi_h^{(R1)}\\
\sqrt{\gamma}\sigma_{hr}^{(Q2)}-\frac{1}{\sqrt2}\beta\Pi_g^{(R1)}%
\end{array}\right],
\begin{array}
[c]{l}%
\frac{\Delta}{2}\sum_{i=1}^3\Pi_r^{(Qi)}+\operatorname{Im}\bigg\{\sqrt{\frac\gamma2}\beta(\sigma_{gr}^{(Q1)*}\Pi_g^{(R1)}-\\
\quad\sigma_{hr}^{(Q2)*}\Pi_g^{(R1)}-\sqrt{2}\sigma_{gr}^{(Q3)*}\Pi_h^{(R1)})\bigg\}%
\end{array}\right),\nonumber\\
G_f^{\prime} &= \left(\frac{1}{\sqrt2}\left[\begin{array}
[c]{lll}%
\Pi_g^{(R2)} & -\Pi_h^{(R2)} & 1\\
-\sqrt{2}\Pi_h^{(R2)} & \sqrt{2}\Pi_g^{(R2)} & 0\\
-\Pi_g^{(R2)} & \Pi_h^{(R2)} & 1%
\end{array}\right],\left[\begin{array}
[c]{l}%
\sqrt{\gamma}\sigma_{gr}^{(Q2)}+\frac{1}{\sqrt2}\beta\Pi_g^{(R2)}\\
\sqrt{\gamma}\sigma_{hr}^{(Q1)}-\beta\Pi_h^{(R2)}\\
\sqrt{\gamma}\sigma_{hr}^{(Q3)}-\frac{1}{\sqrt2}\beta\Pi_g^{(R2)}%
\end{array}\right],
\begin{array}
[c]{l}%
\frac{\Delta}{2}\sum_{i=1}^3\Pi_r^{(Qi)}+\operatorname{Im}\bigg\{\sqrt{\frac\gamma2}\beta(-\sqrt{2}\sigma_{hr}^{(Q1)*}\Pi_h^{(R2)}+\\
\quad\sigma_{gr}^{(Q2)*}\Pi_g^{(R2)}-\sigma_{hr}^{(Q3)*}\Pi_g^{(R2)})\bigg\}%
\end{array}\right).
\end{align}

We now perform the adiabatic elimination procedure on $G_f\boxplus G_f^\prime$ to isolate the Raman dynamics on the cavity qubits.  To do this, we will exploit the trick described in section \ref{sec:Luc's-trick} in order to scale both the input coherent amplitude and excited state detuning, redefining $\beta\rightarrow k\beta$, $\Delta\rightarrow k^2\Delta$, and $G_f\boxplus G_f^\prime\rightarrow G^{(k)}_f\boxplus G_f^{(k)\prime}$ with scaling parameter $k$.  Following the procedure in section \ref{sec:Luc's-trick} we construct the systems
\begin{align}
\tilde{G}_f^{(k)} &= \left(I,\sqrt{\frac\gamma2}\left[\begin{array}
[c]{l}%
\sigma_{gr}^{(Q1)}\Pi_g^{(R1)}-\sqrt{2}\sigma_{gr}^{(Q3)}\Pi_h^{(R1)}-\sigma_{hr}^{(Q2)}\Pi_g^{(R1)}\\
-\sigma_{gr}^{(Q1)}\Pi_h^{(R1)}+\sqrt{2}\sigma_{gr}^{(Q3)}\Pi_g^{(R1)}+\sigma_{hr}^{(Q2)}\Pi_h^{(R1)}\\
\sigma_{gr}^{(Q1)}+\sigma_{hr}^{(Q2)}%
\end{array}\right],
\begin{array}
[c]{l}%
k^2\frac{\Delta}{2}\sum_{i=1}^3\Pi_r^{(Qi)}+2\operatorname{Im}\bigg\{\sqrt{\frac\gamma2}k\beta(\sigma_{gr}^{(Q1)*}\Pi_g^{(R1)}-\\
\quad\sigma_{hr}^{(Q2)*}\Pi_g^{(R1)}-\sqrt{2}\sigma_{gr}^{(Q3)*}\Pi_h^{(R1)})\bigg\}%
\end{array}\right),\nonumber\\
\tilde{G}_f^{(k)\prime} &= \left(I,\sqrt{\frac\gamma2}\left[\begin{array}
[c]{l}%
\sigma_{gr}^{(Q2)}\Pi_g^{(R2)}-\sqrt{2}\sigma_{hr}^{(Q1)}\Pi_h^{(R2)}-\sigma_{hr}^{(Q3)}\Pi_g^{(R2)}\\
-\sigma_{gr}^{(Q2)}\Pi_h^{(R2)}+\sqrt{2}\sigma_{hr}^{(Q1)}\Pi_g^{(R2)}+\sigma_{hr}^{(Q3)}\Pi_h^{(R2)}\\
\sigma_{gr}^{(Q2)}+\sigma_{hr}^{(Q3)}%
\end{array}\right],
\begin{array}
[c]{l}%
k^2\frac{\Delta}{2}\sum_{i=1}^3\Pi_r^{(Qi)}+2\operatorname{Im}\bigg\{\sqrt{\frac\gamma2}k\beta(-\sqrt{2}\sigma_{hr}^{(Q1)*}\Pi_h^{(R2)}+\\
\quad\sigma_{gr}^{(Q2)*}\Pi_g^{(R2)}-\sigma_{hr}^{(Q3)*}\Pi_g^{(R2)})\bigg\}%
\end{array}\right).
\end{align}
Define $\tilde{G}^{(k)}_F = \tilde{G}^{(k)}_f\boxplus \tilde{G}_f^{(k)\prime}$.  Straightforward application of the usual limit theorem \cite{Bout08} taking $k\rightarrow\infty$ yields the limiting system%
\begin{align}\begin{split}
\tilde{G}_F &=\Biggl(I,0,\Omega\biggl(\sqrt{2}X^{(Q1)}\Pi_g^{(R1)}\Pi_h^{(R2)}+X^{(Q2)}\Pi_g^{(R1)}\Pi_g^{(R2)}-\sqrt{2}X^{(Q3)}\Pi_h^{(R1)}\Pi_g^{(R2)}-\\
&\quad\quad\quad\quad\Pi_g^{(R1)}(\Pi_g^{(Q1)}+\Pi_h^{(Q2)})-2\Pi_h^{(R1)}\Pi_g^{(Q3)}-\Pi_g^{(R2)}(\Pi_g^{(Q2)}+\Pi_h^{(Q3)})-2\Pi_h^{(R2)}\Pi_h^{(Q1)}
\biggr)\Biggr)
\end{split}\end{align}
where the Raman interaction strength is now set by $\Omega = \gamma\vert\beta\vert^2/2\Delta$.  
Again following the procedure of section \ref{sec:Luc's-trick}, we construct $G_F^{(k)}$, the asymptotic approximation of  $G^{(k)}_f\boxplus G_f^{(k)\prime}$ for sufficiently large values of $k$.
\begin{align}\begin{split}
G_F^{(k)} &= \Biggl(\text{diag}\left\{S^{(f)},S^{\prime(f)}\right\},\left[\begin{array}
{c}%
\frac{1}{\sqrt2}k\beta\Pi_g^{(R1)}\\
-k\beta\Pi_h^{(R1)}\\
-\frac{1}{\sqrt2}k\beta\Pi_g^{(R1)}\\
\frac{1}{\sqrt2}k\beta\Pi_g^{(R2)}\\
-k\beta\Pi_h^{(R2)}\\
-\frac{1}{\sqrt2}k\beta\Pi_g^{(R2)}
\end{array}\right],\Omega\biggl(\sqrt{2}X^{(Q1)}\Pi_g^{(R1)}\Pi_h^{(R2)}+X^{(Q2)}\Pi_g^{(R1)}\Pi_g^{(R2)}-\sqrt{2}X^{(Q3)}\Pi_h^{(R1)}\Pi_g^{(R2)}-\\
&\quad\quad\quad\quad\Pi_g^{(R1)}(\Pi_g^{(Q1)}+\Pi_h^{(Q2)})-2\Pi_h^{(R1)}\Pi_g^{(Q3)}-\Pi_g^{(R2)}(\Pi_g^{(Q2)}+\Pi_h^{(Q3)})-2\Pi_h^{(R2)}\Pi_h^{(Q1)}
\biggr)\Biggr).
\end{split}\end{align}
The analogous phase-flip network simply replaces $X^{(Qi)}$ with $Z^{(Qi)}$  and $2\Pi^{(Qi)}_{^h_g}\rightarrow I\pm X^{(Qi)}$ in the Hamiltonian above.

\subsubsection{Network master equation}\label{sec:networkME}
Concatenating the asymptotically approximate `correction' subsystems calculated above with the error processes, we immediately arrive at a master equation for the bit-flip network (see section \ref{sec:master-equation}):
\begin{align}\label{eq:networkME}
\frac{d\rho_{t}}{dt}  & =-i\left[  H,\rho_{t}\right]  +\sum_{j=1}^n\left\{  L_{j}\rho_{t}%
L_{j}^{\,\ast}-\frac{1}{2}L_{j}^{\,\ast}L_{j}\rho_{t}-\frac{1}{2}\rho_{t}L_{j}^{\,\ast}L_{j}\right\};
\end{align}
\begin{align}\label{eq:networkCoef}
H &=\Omega\biggl(\sqrt{2}X^{(Q1)}\Pi_g^{(R1)}\Pi_h^{(R2)}+X^{(Q2)}\Pi_g^{(R1)}\Pi_g^{(R2)}-\sqrt{2}X^{(Q3)}\Pi_h^{(R1)}\Pi_g^{(R2)}-\nonumber\\
&\quad\quad\quad\quad\Pi_g^{(R1)}(\Pi_g^{(Q1)}+\Pi_h^{(Q2)})-2\Pi_h^{(R1)}\Pi_g^{(Q3)}-\Pi_g^{(R2)}(\Pi_g^{(Q2)}+\Pi_h^{(Q3)})-2\Pi_h^{(R2)}\Pi_h^{(Q1)}
\biggr),\nonumber\\
L_1 &= \frac{\alpha}{\sqrt2}\left(\Pi_g^{(R1)}O^{(12)}+\sigma_{hg}^{(R1)}E^{(12)}\right),\nonumber\\
L_2 & = \frac{\alpha}{\sqrt2}\left(-\sigma_{gh}^{(R1)}O^{(12)}-\Pi_h^{(R1)}E^{(12)}\right),\nonumber\\
L_3 &= \frac{\alpha}{\sqrt2}\left(\Pi_g^{(R2)}O^{(32)}+\sigma_{hg}^{(R2)}E^{(32)}\right),\nonumber\\
L_4 & = \frac{\alpha}{\sqrt2}\left(-\sigma_{gh}^{(R2)}O^{(32)}-\Pi_h^{(R2)}E^{(32)}\right),\nonumber\\
L_5 &= \sqrt{\Gamma}X^{(Q1)},\quad L_6 = \sqrt{\Gamma}X^{(Q2)},\quad L_7 = \sqrt{\Gamma}X^{(Q3)}.
\end{align}
The phase-flip network simply replaces $X^{(Qi)}\leftrightarrow Z^{(Qi)}$ and $2\Pi^{(Qi)}_{^h_g}\rightarrow I\pm X^{(Qi)}$ in the above.  Note that we have omitted the field coupling operators from $G_F^{(k)}$ in the above master equation.  This is because these terms serve only to decohere the ground states of the relays, which have no effect on our network dynamics if the relays are initiated in ground state eigenstates.  The Hamiltonian dynamics principally serve to flip the state of the individual qubits, depending on the state of the relays (first three terms); $L_1$ and $L_2$ drive the state of R1 to reflect the joint parity of Q1 and Q2, and $L_3$ and $L_4$ drive R2 to reflect the parity of Q3 and Q2.  The final four terms of the Hamiltonian are physically recognizable as AC Stark shifts that necessarily accompany the Raman interactions.  Although not immediately obvious, the feedback was designed to minimize the detrimental effects of these shifts, as described in section \ref{sec:SS}.

\subsection{AC Stark shift compensation}\label{sec:SS}
 
\begin{table}
\begin{tabular}{c|cc|ccc}
\hline
Q1,Q2,Q3 & R1 & R2 & SS1 & SS2 & SS3\\
\hline
\hline
h,h,h & h & h & 0 & 0 & -2$\Omega$\\
g,g,g & h & h & -2$\Omega$ & 0 & 0\\
\hline
h,g,g & g & h & -2$\Omega$ & 0 & 0\\ 
g,h,h & g & h & -1$\Omega$ & -1$\Omega$ & 0\\
\hline
h,g,h & g & g & -1$\Omega$ & -1$\Omega$ & 0\\
g,h,g & g & g &  0 & -1$\Omega$ & -1$\Omega$\\
\hline
h,h,g & h & g & 0 & -1$\Omega$ & -1$\Omega$\\
g,g,h & h & g & 0 & 0 & -2$\Omega$\\ 
\hline
\end{tabular}\caption{{\bf AC Stark Shifts of error states}  AC Stark shifts for each three-qubit basis state in the bit-flip network described in section \ref{sec:net-calculations}.  According to the Hamiltonian in equation \eqref{eq:networkCoef}, for joint each qubit state (Q1,Q2,Q3) and corresponding relay states that correctly reflect its parities (R1,R2), the AC Stark shift experienced by each qubit is listed under the columns (SS1, SS2, SS3).  As designed, each joint qubit state will experience an aggregate $-2\Omega$ AC Stark shift, as long as the error state is correctly represented in the state of the relays.}\label{tab:SS}
\end{table}

As discussed in sections \ref{sec:Raman-derivation} and \ref{sec:networkME}, our feedback network induces both Raman interactions that help to correct errors and AC Stark shifts that alter the energy of each qubit ground state.  Unfortunately, these Stark shifts tend to break the energy degeneracy between joint qubit states, which consequently causes rapid dephasing in the encoded codeword.  There are two ways to counteract this.  One is to \emph{design the network} such that the net effect of these shifts is null and the other is to \emph{modify the components} such that the AC Stark shift is nulled on a qubit-by-qubit basis.

The network described in section \ref{sec:net-calculations} takes the former approach.  As summarized in table \ref{tab:SS}, by choosing the intensities of and the particular transitions targeted by the Raman beams, the feedback is designed such that the aggregate AC Stark shift experienced by any joint qubit state is equal to $-2\Omega$, as long as the relays accurately reflect the joint parities of the qubits.  Thus, codeword dephasing occurs only in the interval between changes in the qubit parities and the accurate reflection of these changes in the relays.  Consequently, the amount of Stark shift-induced dephasing may be decreased by increasing the probe amplitude $\alpha$.  This approach yields sub-optimal storage fidelities, as some codeword dephasing will occur for finite $\alpha$, but has the advantage of significantly simpler components.  Representative performance curves are shown in Fig.~\ref{fig:meq}.

\begin{figure}[tb!]
\includegraphics[width=0.45\textwidth]{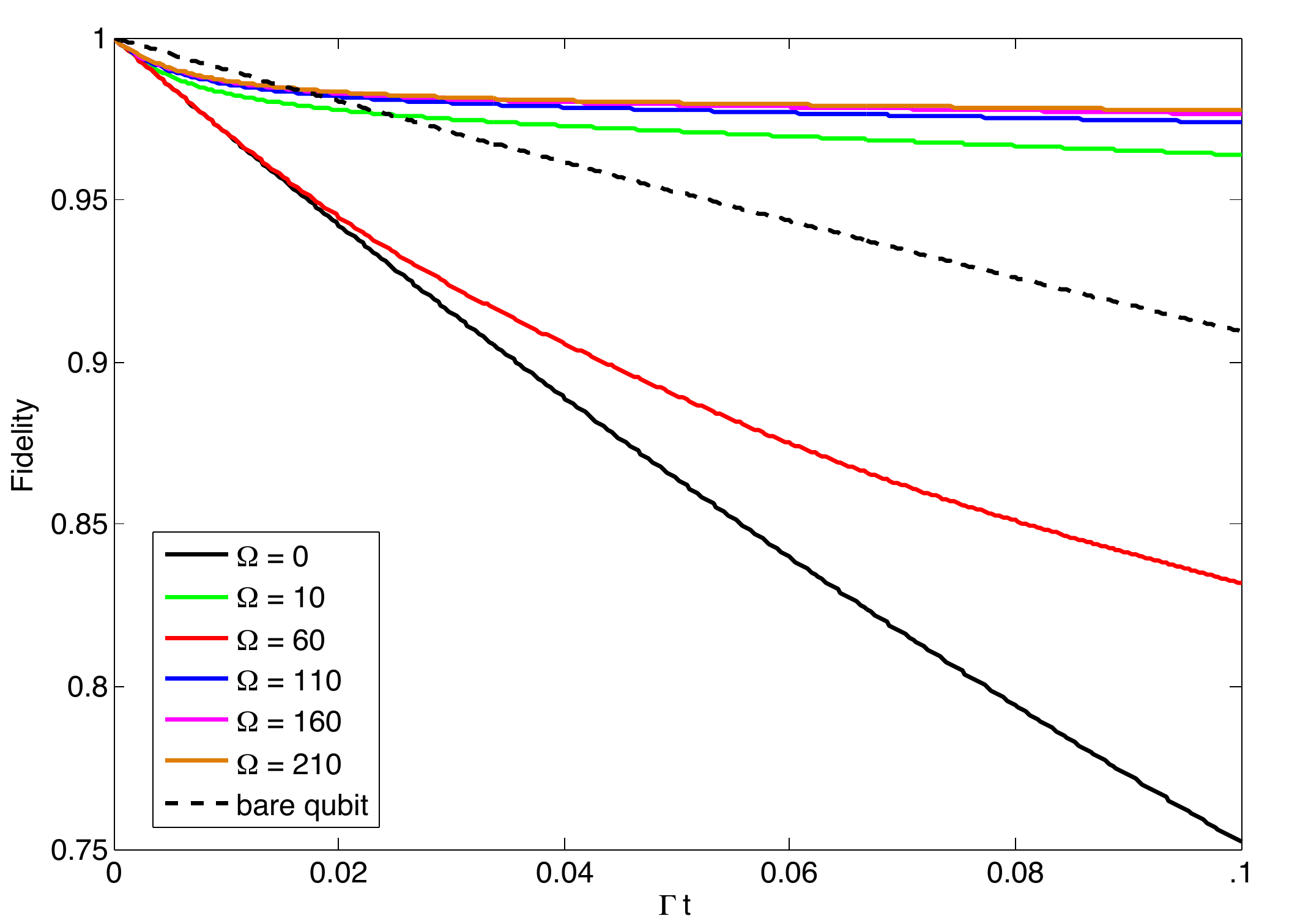}
\caption{\label{fig:meq_SScomp} {\bf Performance with AC Stark shift compensation.}  Decay of fidelity, $\langle\Psi_0\vert \rho_t\vert\Psi_0\rangle$, for  several values of the feedback parameter $\Omega=\beta^2\gamma/2\Delta$, $\alpha=\Omega/8$, and $\Gamma=0.1$ when extra, AC Stark shift compensating qubit levels are included.  The fidelity decay of a single, bare qubit also suffering bitflip errors at rate $\Gamma=.1$ is also shown.}
\vspace{-0.1in}
\end{figure}

The latter approach was discussed in some detail at the end of section \ref{sec:Raman-derivation}.  By presuming an extra pair of atomic excited states in each qubit with equal but opposite detuning from the Raman beams, the AC Stark shifts may be completely cancelled on a qubit-by-qubit basis. (Alternatively, an additional pair of feedback Raman beams with equal but opposite detunings would have an equivalent effect)  The resulting network master equation would have an identical form as equations \eqref{eq:networkME} and \eqref{eq:networkCoef}, but without the final four terms in the Hamiltonian.  Such a network yields better storage fidelities than the network described in section \ref{sec:net-calculations}, but requires significantly more complicated and precisely tuned component devices.  Representative performance curves produced by this approach are depicted in Fig. \ref{fig:meq_SScomp}, which were calculated for the same parameters as for Fig.~\ref{fig:meq}.

\end{document}